\documentclass[letterpaper,twocolumn,10pt]{article}
\RequirePackage[hyphens]{url}
\usepackage{usenix2019_v3}
\usepackage[total={7in,9in},top=1in,left=0.75in,right=0.75in,bottom=1in,headsep=0.5in]{geometry}
\usepackage{mathptmx}
\usepackage{paralist}
\usepackage{xspace}
\usepackage{cancel}
\usepackage{pifont}
\usepackage{comment}
\usepackage{listings}
\usepackage{wrapfig}
\usepackage{bm}
\usepackage[font=small]{caption}
\usepackage[frozencache]{minted}
\usepackage{subcaption}
\DeclareCaptionSubType{listing}
\usepackage{multirow}
\usepackage{booktabs}
\usepackage[boxed,lined,noend]{algorithm2e}
\usepackage{enumitem}
\usepackage{paralist}
\usepackage{fancyhdr}
\usepackage[normalem]{ulem}
\usepackage{titlesec}
\usepackage{tikz}
\usepackage{cleveref}
\usepackage{graphicx}

\hypersetup{bookmarks=false,pdfstartpage=1,pdfpagemode=UseNone,
            colorlinks,linkcolor={red!50!black},
            citecolor={blue!50!black},urlcolor={blue!80!black}}

\def\BibTeX{{\rm B\kern-.05em{\sc i\kern-.025em b}\kern-.08em
    T\kern-.1667em\lower.7ex\hbox{E}\kern-.125emX}}

\makeatletter
\renewcommand{\paragraph}{%
  \@startsection{paragraph}{4}%
  {\z@}{1ex \@plus .2ex \@minus .2ex}{-1em}%
  {\normalfont\normalsize\bfseries}%
}
\makeatother

\newcommand*\circled[1]{\tikz[baseline=(char.base)]{
            \node[shape=circle,draw,inner sep=1pt] (char) {#1};}}

\makeatletter
\newcommand{\crefnames}[3]{%
  \@for\next:=#1\do{%
    \expandafter\crefname\expandafter{\next}{#2}{#3}%
  }%
}
\makeatother
\crefnames{part,chapter,section}{\S}{\S\S}
\crefnames{para,paragraph,subparagraph}{\P}{\P\P}
\crefnames{listing}{Listing}{Listings}
\crefnames{sublisting}{Listing}{Listings}
\crefnames{figure}{Figure}{Figures}
\crefnames{table}{Table}{Tables}
\crefnames{appendix}{Appendix}{Appendices}

\definecolor{attackRed}{RGB}{228,72,56}

\definecolor{backgroundColor}{rgb}{0.95,0.95,0.92}
\setminted[gas]{
  frame=single,
  framesep=2mm,
  fontsize=\footnotesize,
  escapeinside=||,
  numbersep=5pt,
  mathescape,
}
\setminted[python]{
  frame=single,
  framesep=2mm,
  fontsize=\footnotesize,
  escapeinside=||,
  mathescape,
}
\setminted[c]{
  xleftmargin=\parindent,
  numbersep=5pt,
  frame=single,
  framesep=2mm,
  baselinestretch=1.2,
  fontsize=\footnotesize,
  escapeinside=||,
}

\makeatletter
\renewcommand{\section}{\@startsection{section}{1}{0pt}%
{-3ex plus -1ex minus -.2ex}{1.5ex plus.2ex}%
{\reset@font\large\bf}}
\renewcommand{\subsection}{\@startsection{subsection}{1}{0pt}%
{-2ex plus -1ex minus -.2ex}{1ex plus.2ex}%
{\reset@font\large\bf}}
\makeatother

\begin{document}

\date{}

\title{\Large \bf An Analysis of Speculative Type Confusion Vulnerabilities in the Wild}

\author{
{\rm Ofek Kirzner \qquad Adam Morrison}\\
Tel Aviv University
}

\maketitle
\pagestyle{empty}

\begin{abstract}
Spectre v1 attacks, which exploit conditional branch misprediction, are often identified
with attacks that bypass array bounds checking to leak data from a victim's memory.
Generally, however, Spectre v1 attacks can exploit \emph{any} conditional branch misprediction that
makes the victim execute code incorrectly.  In this paper, we investigate
\emph{speculative type confusion}, a Spectre v1 attack vector in which branch mispredictions
make the victim execute with variables holding values of the wrong type and thereby leak memory content.

We observe that speculative type confusion can be inadvertently introduced by a compiler,
making it extremely hard for programmers to reason about security and manually apply Spectre mitigations.
We thus set out to determine the extent to which speculative type confusion affects the Linux
kernel.  Our analysis finds exploitable and potentially-exploitable
\emph{arbitrary} memory disclosure vulnerabilities.  We also find many \emph{latent} vulnerabilities,
which could become exploitable due to innocuous system changes, such as coding style changes.

Our results suggest that Spectre mitigations which rely on statically/manually identifying
``bad'' code patterns need to be rethought, and more comprehensive mitigations are needed.
\end{abstract}

\section{Introduction} \label{sec:introduction}

Spectre attacks~\cite{spectre,spec_variant_four,speculative_buffer_overflow,spectre_returns,ret2spec,SpecROP,Blindside}
exploit processor control- or data-flow speculation to leak data from a victim program's memory over
a microarchitectural covert channel. A Spectre attack maneuvers the processor to mispredict the
correct execution path of an instruction sequence in the victim, referred to as a \emph{Spectre gadget}.
The gadget's misspeculated execution acts as a ``confused deputy'' and accesses data from an
attacker-determined location in the victim's address space.
Although the mispredicted instructions are \emph{transient}~\cite{spectre,sok_spectre_meltdown}---%
the CPU eventually discards them without committing their results to architectural state---%
data-dependent traces of their execution remain observable in microarchitectural state, such as the cache.
These data-dependent side effects form a covert channel that leaks the accessed data to the
attacker.

Spectre attacks pose a serious threat to monolithic operating system (OS) kernels.  While Spectre
attacks can only leak data architecturally accessible to the victim, a victim that is an OS kernel
typically has all physical memory mapped in its virtual address space and thus architecturally accessible~\cite{linux-kernel-map,windows-kernel-map}. Moreover, kernels expose a large attack surface (e.g., hundreds of system calls) through which
an attacker can trigger Spectre gadgets.

Since speculative execution is fundamental to modern processor design, processor vendors do not plan to mitigate
Spectre attacks completely in hardware~\cite{intel-spectre-v1-fix,intel-spectre-v1-ceo,amd-spectre-v1-fix}.%
\footnote{In contrast, Meltdown-type attacks~\cite{meltdown,ridl,Schwarz2019ZombieLoad,foreshadow,lvi}
exploit an Intel processor implementation artifact (addressed in future processors~\cite{intel-spectre-v1-ceo}),
wherein a load targeting \emph{architecturally inaccessible} data (e.g., in another address space)
can execute before being discarded due to a virtual memory exception.}
Vendors instead suggest using software mitigations to restrict speculation~\cite{intel-spectre-v1-fix,amd-spectre-v1-fix}.
To minimize their performance impact, most software mitigations target specific ``templates'' of potentially
vulnerable gadgets, which are identified with static or manual analysis~\cite{msvc-mitigations,oo7,linux-mitigations,ICC-mitigation}.%
\footnote{We discuss more comprehensive software mitigations---which, unfortunately, have high performance overheads---in~\cref{sec:discussion}.}

\begin{listing}[t]
\captionsetup[sublisting]{aboveskip=-2pt}
\begin{sublisting}{.5\textwidth}
\begin{minted}{c}
if (|\colorbox{attackRed}{\textbf{x}}| < array1_len) { // branch mispredict: taken
  y = array1[|\colorbox{attackRed}{\textbf{x}}|];      // read out of bounds
  z = array2[y * 4096]; } // leak y over cache channel
\end{minted}
\caption{Bounds check bypass.}
\label{lst:spectre-example-bypass}
\end{sublisting}\\
\begin{sublisting}{.5\textwidth}
\begin{minted}{c}
void syscall_helper(cmd_t* cmd, char* ptr, long |\colorbox{attackRed}{\textbf{x}}|) {
  // ptr argument held in x86 register %rsi
  cmd_t c = *cmd;   // cache miss
  if (c == CMD_A) { // branch mispredict: taken
    ... code during which |\colorbox{attackRed}{\textcolor{black}{\textbf{x}}}| moves to %rsi ...
  }
  if (c == CMD_B) { // branch mispredict: taken
    y = *ptr;  // read from addr |\colorbox{attackRed}{\textcolor{black}{\textbf{x}}}| (now in |\colorbox{attackRed}{\textcolor{black}{\textbf{%rsi}}}|)
    z = array[y * 4096]; // leak y over cache channel
  }
  ... rest of function ...
\end{minted}
\caption{Type confusion.}
\label{lst:spectre-example-tc}
\end{sublisting}
\caption{Spectre gadgets for exploiting conditional branch prediction.  Data in \colorbox{attackRed}{\textbf{red boxes}} is attacker-controlled.}
\label{lst:spectre-example}
\end{listing}

In this paper, we focus on conditional branch prediction Spectre attacks (so-called ``variant 1''~\cite{spectre}).
These attacks are often characterized as \emph{bounds check bypass} attacks, which exploit misprediction of an array
bounds check to perform an out-of-bounds access and leak its result~(\cref{lst:spectre-example-bypass}).
Deployed software mitigations in compilers and OS kernels target this type of gadget template~\cite{msvc-mitigations,linux-mitigations,ICC-mitigation}.

Generally, however, a Spectre attack is defined as exploiting conditional branch prediction to make the processor
\emph{``temporarily violate program semantics by executing code that would not have been executed otherwise''}~\cite{spectre}---and
a bounds check bypass is just one example of such a violation.  Speculative \emph{type confusion} is a different
violation, in which misspeculation makes the victim execute with some variables holding values of the
wrong type, and thereby leak memory content.

\Cref{lst:spectre-example-tc} shows an example \emph{compiler-introduced} speculative type confusion gadget, which causes the victim to dereference
an attacker-controlled value. In this example, the compiler emits code for the first \texttt{if} block that clobbers
the register holding a (trusted) pointer with an untrusted value, based on the reasoning that if the first \texttt{if} block
executes, then the second \texttt{if} block will not execute. Thus, if the branches mispredict such that both blocks
execute, the code in the second \texttt{if} block leaks the contents of the attacker-determined location.  In contrast to
the bounds check bypass attack, here the attacker-controlled address has no data-dependency on the branch predicate, nor does
the predicate depend on untrusted data.  Consequently, \emph{this gadget would not be protected by existing OS kernel Spectre mitigations,
nor would programmers expect it to require Spectre protection}.

To date, speculative type confusion has mainly been hypothesized about, and even then, only in the context of object-oriented polymorphic
code~\cite{sok_spectre_meltdown} or as a vector for bypassing bounds checks~\cite{MSFT_stc,javascript-stc}.  Our key driving observation
in this paper is that speculative type confusion may be much more prevalent---as evidenced, for instance, by~\cref{lst:spectre-example-tc},
which does not involve polymorphism or bounds checking.  Accordingly, we set out to answer the following question: \emph{are OS kernels
vulnerable to speculative type confusion attacks?}

\subsection{Overview \& contributions}

We study Linux,
which dominates the OS market share for server and mobile computers~\cite{os-market-share}.
In a nutshell, not only do we find exploitable speculative type confusion vulnerabilities, but---perhaps more disturbingly---our analysis indicates that OS kernel security currently rests on shaky foundations.
There are many latent vulnerabilities that are not exploitable only due to serendipitous circumstances, and may be rendered exploitable by
different compiler versions, innocuous code changes, deeper-speculating future processors, and so on.%
\footnote{Indeed, we make no security claims for these ``near miss'' vulnerabilities; some of them may be exploitable in
kernel versions or platforms that our analysis---which is not exhaustive---does not cover.}

\paragraph{Attacker-introduced vulnerabilities (\cref{sec:ebpf})}
Linux supports untrusted user-defined kernel extensions, which are loaded in the form of eBPF%
\footnote{Extended Berkeley Packet Filter~\cite{eBPF,BPF}.}
bytecode programs.  The kernel verifies the safety of extensions using static analysis and compiles them to native code
that runs in privileged context.  The eBPF verifier does not reason about speculative control flow, and thus successfully
verifies eBPF programs with speculative type confusion gadgets.  eBPF emits Spectre mitigations into
the compiled code, but these only target bounds check bypass gadgets.  Consequently, we demonstrate that an unprivileged user
can exploit eBPF to create a Spectre \emph{universal read gadget}~\cite{spectre_google} and read arbitrary physical memory
contents at a rate of 6.7\,KB/sec with $99\%$ accuracy.

\paragraph{Compiler-introduced vulnerabilities (\cref{sec:compilers})}
We show that C compilers can emit speculative type confusion gadgets.
While the gadgets are blocked by full Spectre compiler mitigation modes (e.g., speculative load hardening (SLH)~\cite{SLH}),
these modes have high performance overheads (\cref{sec:discussion}), and in GCC must be manually enabled per-variable.
Optional low-overhead mitigation modes in Microsoft and Intel compilers do not block these gadgets.
Motivated by these findings, we perform a binary-level analysis of Linux
to determine whether it contains
speculative type confusion introduced by compiler optimizations.
We find several such cases. In assessing potential exploitability of these cases, we investigate how x86 processors resolve
mispredicted branches. We find that Spectre gadgets which today may be considered unexploitable are actually exploitable,
which may be of independent interest.

\paragraph{Polymorphism-related vulnerabilities (\cref{sec:jumpswitch})}
The Linux kernel makes heavy use of object-oriented techniques, such as data inheritance and polymorphism,
for implementing kernel subsystem interfaces.  The related indirect function calls are protected with retpolines~\cite{retpoline},
which essentially disable indirect call prediction.  To claw back the resulting lost performance, Linux replaces
certain indirect calls with direct calls to one of a set of legal call targets, where the correct target is chosen
using conditional branches~\cite{retpoline-pain,234862}.  Unfortunately, this approach opens the door to speculative type
confusion among the different targets implementing a kernel interface.  We perform a source-level analysis on Linux to
find such vulnerabilities.  We identify dozens of \emph{latent} vulnerabilities, namely: vulnerable gadgets which are
not exploitable by chance, and could become exploitable by accident. For example, we find gadgets in which the attacker
controls a 32-bit value, which cannot represent a kernel pointer on 64-bit machines.  But if a future kernel version
makes some variables 64-bit wide, such gadgets would become exploitable.

\subsection{Implications}
Our work shows that speculative type confusion vulnerabilities are more insidious than speculative bounds check bypasses,
with exploitable and latent vulnerabilities existing in kernel code.  Given the existence of compiler-introduced
vulnerabilities, we question the feasibility of the current Linux and GCC mitigation approach, which relies on developers
manually protecting ``sensitive'' variables~\cite{linux-mitigations}, likely due to equating Spectre v1 with bounds check bypasses.
While comprehensive mitigations, such as SLH~\cite{SLH}, can block all Spectre attacks, they impose significant overhead on kernel operations (up to $2.7\times$, see~\cref{sec:discussion}). It is also unclear whether static analysis~\cite{Spectector,ConstTimeFoundations} can be
incorporated into the OS kernel development process to guarantee absence of speculative type confusion vulnerabilities.
In short, current Spectre mitigations in OS kernels
require rethinking and further research.

\section{Background} \label{sec:background}

\subsection{Out-of-order \& speculative execution} \label{sec:spec-exec}

Modern processors derive most of their performance from two underlying mechanisms: out-of-order (OoO)
execution~\cite{tomasulo1967efficient} and speculation~\cite{hennessy2011computer}.

\paragraph{OoO execution}
A processor core consists of a \emph{frontend}, which fetches instruction from memory, and a \emph{backend},
responsible for instruction execution.  A fetched instruction is \emph{decoded} into internal micro-operations ($\mu$-ops),
which are then \emph{dispatched} to a \emph{reservation station} and await execution.  Once the operands of a $\mu$-op
become available (i.e., have been computed), it is \emph{issued} to an execution unit
where it is \emph{executed}, making its result available to dependent $\mu$-ops.  Hence, $\mu$-ops may execute
out of program order.  To maintain the program order and handle exceptions, $\mu$-ops are queued into a reorder buffer
(ROB) in program order.  Once a $\mu$-op reaches the ROB head and has been executed, it gets \emph{retired}:
its results are committed to architectural state and any pipeline resources allocated to it are freed.

\paragraph{Speculative execution}
To execute instructions as soon as possible, the processor attempts to \emph{predict} the results of certain
(usually long latency) $\mu$-ops.  The prediction is made available to dependent $\mu$-ops, allowing them
to execute.  Once the predicted $\mu$-op executes, the backend checks if the $\mu$-op's output was correctly predicted.
If so, the $\mu$-op proceeds to retirement; otherwise, the backend \emph{squashes} all $\mu$-ops following the
mispredicted $\mu$-op in the ROB and reverts its state to the last known correct state (which was checkpointed when
the prediction was made).
We refer to the maximum amount of work that can be performed in the shadow of a
speculative $\mu$-op as the \emph{speculation window}. It is determined by the latency of computing the predicted
$\mu$-op's results and the available microarchitectural resources (e.g., the size of the ROB limits how many
$\mu$-ops can be in flight).
We consider \emph{control-flow speculation}, described in the following section.

\subsection{Branch prediction} \label{sec:BPU}

To maximize instruction throughput, the processor performs branch prediction in the frontend.
When a branch is fetched, a \emph{branch predictor unit} (BPU) predicts its outcome, so that the frontend can continue
fetching instructions from the (predicted) execution path without stalling.  The branch is \emph{resolved}
when it gets executed and the prediction is verified, possibly resulting in a squash and re-steering of
the frontend.  Notice that \emph{every} branch is predicted, even if its operands are readily available
(e.g., in architectural state), because figuring out availability of operands is only done in the backend.

\begin{figure}
	\includegraphics[width=\linewidth]{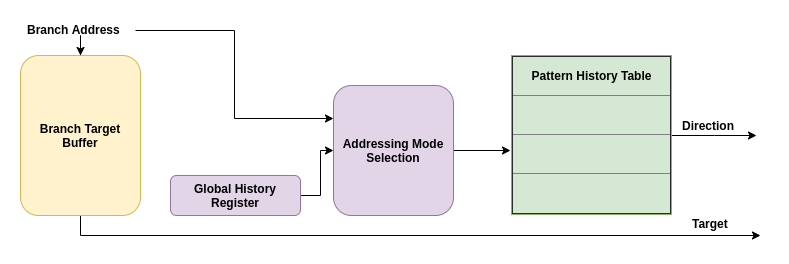}
	\caption{BPU Scheme}
	\label{fig:bpu}
    \vspace{-10pt}
\end{figure}

We assume the branch predictor unit design shown in~\cref{fig:bpu}, which appears to match Intel's
BPU~\cite{BranchScope, BPforTrojans}.  The BPU has two main components: an outcome predictor, predicting the
direction of conditional branches, and a \emph{branch target buffer} (BTB), predicting the target address of
indirect branches.
The outcome predictor stores 2-bit saturating counters in a \emph{pattern history table} (PHT).  A branch's
outcome is predicted based on a PHT entry selected by hashing its program counter (PC).  The PHT entry is
selected in one of two addressing modes, depending on the prediction success rate:  1- or 2-level prediction.
The 1-level predictor uses only the PC, whereas the 2-level predictor additionally hashes a \emph{global history
register} (GHR) that records the outcome of the previously encountered branches.

\subsection{Cache covert channels}

To hide memory latency, the processor contains fast memory buffers called \emph{caches}, which hold
recently and frequently accessed data.  Modern caches are \emph{set-associative}: the cache is organized into multiple
\emph{sets}, each of which can store a number of cache lines.
The cache looks up an address by hashing it to obtain a set, and then searching all cache lines in
that set.

Changes in cache state can be used to construct a covert channel.  Consider transmission of a $B$-bit symbol $x$.
In a {\sc Flush+Reload} channel~\cite{flush+}, (1) the receiver flushes $2^B$ lines from the cache; (2) the sender accesses the $x$-th
line, bringing it back into the cache; and (3) the receiver measures the time it takes to access each line,
identifying $x$ as the only cache hit. {\sc Flush+Reload} requires the sender and receiver to share memory.
A {\sc Prime+Probe} channel avoids this requirement by using evictions instead of line fills~\cite{last_level_cache_practical}.

\subsection{Transient execution attacks}

Transient execution attacks overcome architectural consistency by using the microarchitectural traces left by
\emph{transient}---doomed to squash---$\mu$-ops to leak architecturally-inaccessible data to the attacker.
Meltdown-type attacks~\cite{sok_spectre_meltdown,meltdown,ridl,Schwarz2019ZombieLoad,foreshadow} extract data
across architectural protection domains by exploiting transient execution after a hardware exception.
Our focus, however, is on Spectre-type attacks~\cite{spectre,spec_variant_four,speculative_buffer_overflow,spectre_returns,ret2spec,SpecROP,Blindside},
which exploit misprediction.  Spectre-type attacks maneuver the victim into leaking its own data, and are
limited to the depth of the speculation window.
Spectre v1 (Spectre-PHT~\cite{sok_spectre_meltdown}) exploits misprediction of
conditional branch outcomes. Spectre v2 (Spectre-BTB~\cite{sok_spectre_meltdown}) exploit misprediction of
indirect branch target addresses. The original Spectre v2 attacks poisoned the BTB to redirect control-flow to arbitrary
locations; but as we shall see, even mispredicting a legal target is dangerous (\cref{sec:jumpswitch}). Other variants target
return address speculation~\cite{spectre_returns,ret2spec} or data speculation~\cite{spec_variant_four}.

\section{Threat model} \label{sec:threat}

We consider an attacker who is an unprivileged (non-root)
user on a multi-core machine running the latest Linux kernel.
The attacker's goal is to obtain information located in
the kernel address space, which is not otherwise accessible to it.

We assume the system has all state-of-the-art mitigations against
transient execution attacks enabled.  In particular, the attacker
cannot mount cross-protection domain Meltdown-type attacks to directly
read from the kernel address space.

We assume that the attacker knows kernel virtual addresses.  This knowledge can
be obtained by breaking kernel address space layout randomization (KASLR) using
orthogonal side-channel attacks~\cite{KASLR-break,cacheout,MaliciousMMU} or even simply
from accidental information leak bugs in the kernel.

\section{Speculative type confusion in eBPF} \label{sec:ebpf}

Linux's extended Berkeley Packet Filter (eBPF)~\cite{eBPF} subsystem strives to let the Linux kernel safely
execute untrusted, user-supplied kernel extensions in privileged context.
An eBPF extension is ``attached'' to designated kernel events, such as system call execution or packet processing,
and is executed when these events occur.
eBPF thereby enables applications to customize the kernel for performance monitoring~\cite{iovisor},
packet processing~\cite{XDP}, security sandboxing~\cite{seccomp},~etc.

Unprivileged users can load eBPF extensions into the kernel as of Linux v4.4 (2015)~\cite{bpf-accessible-to-non-priv}.%
\footnote{An \texttt{unprivileged\_bpf\_disable} configuration knob exists for disallowing unprivileged
access to eBPF; it is not set by default.}
An eBPF extension is loaded in the form of a bytecode program for an in-kernel virtual machine (VM),
which is limited in how it can interact with the rest of the kernel.  The kernel statically verifies
the safety of loaded eBPF programs.  On the x86-64 and Arm architectures, the kernel compiles eBPF programs to
native code; on other architectures, they run interpreted.

\begin{figure}
	\includegraphics[scale=0.4]{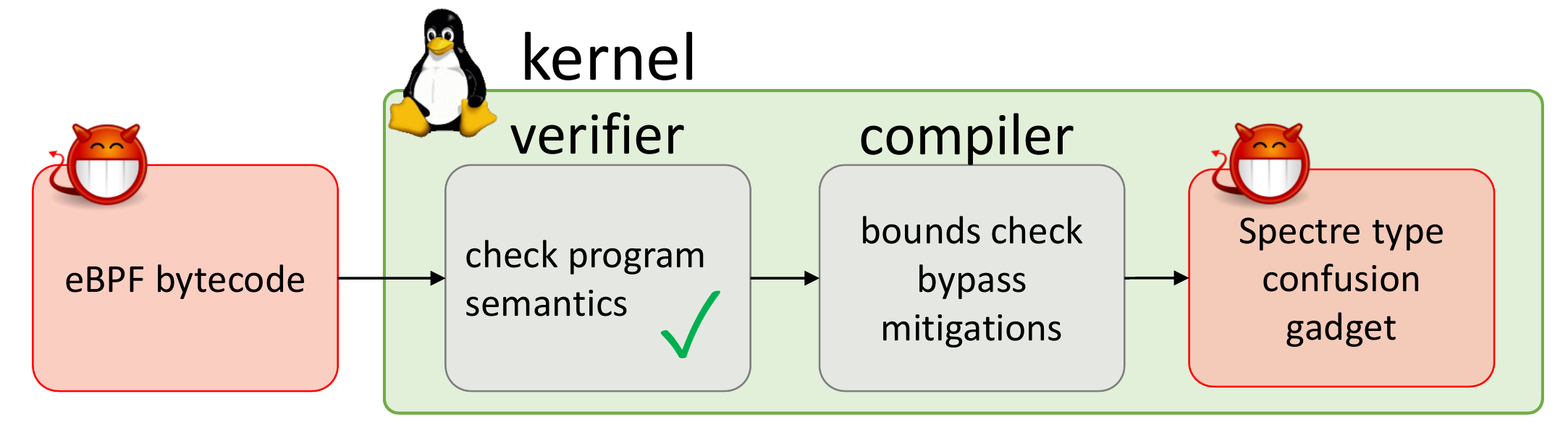}
	\caption{eBPF speculative type confusion attack.}
	\label{fig:ebpf-attack}
\end{figure}

Both eBPF verification and compilation do not consider speculative type confusion, which allows
an attacker to load eBPF programs containing Spectre gadgets and thus read from anywhere in the
kernel address space (\cref{fig:ebpf-attack}).
In the following, we describe the eBPF security model (\cref{sec:ebpf-verifier}), detail its
vulnerability (\cref{sec:ebpf-vuln}), and describe our proof-of-concept attack (\cref{sec:ebpf-exploit}).

\subsection{eBPF security model} \label{sec:ebpf-verifier}

In general, eBPF does not enforce safety at run time, but by statically verifying that the eBPF
program maintains memory safety and is otherwise well-behaved.  (One exception to this principle
are Spectre mitigations, discussed below.)  An eBPF program can only call one of a fixed set
of \emph{helper} functions in the kernel.  The allowed set of helpers depends on the type of eBPF
program and on the privileges of the user that loaded it.

An eBPF program is restricted to accessing a fixed set of memory regions, known at compile time,
including a fixed-size stack and a \emph{context}, which is a program-specific object type that stores
the program's arguments.  An eBPF program can further access statically-sized key/value dictionaries
called \emph{maps}.  The size of a map's keys and values is fixed at map creation time.  A map's
data representation (array, hash table, etc.) is similarly specified on creation.
Maps can be shared by different eBPF programs and can also be accessed by user processes.
An eBPF program that successfully looks up a value in a map receives a pointer to the value, and so
can manipulate it in-place.

\paragraph{eBPF verification}
The kernel statically verifies the safety of an eBPF program in two steps.  The first step performs
control-flow validation, to verify that the program (which runs in privileged context) is guaranteed
to terminate in a fixed amount of time.  This property is verified by checking that the program does
not contain loops and does not exceed a certain fixed size.  The second step verifies the program's
memory safety.  Memory safety is verified by enumerating every possible execution flow to prove
that every memory access is safe.  When processing a flow, the verifier maintains a \emph{type} for
each register and stack slot, and checks that memory accesses are valid with respect to these types.

The verifier tracks whether each register is uninitialized, holds a scalar (non-pointer) value, or
holds a pointer.  Pointers are further typed according to the region they point to: the stack, the context,
a map, a value from a map, and so on. The verifier also tracks whether a pointer is non-NULL
and maintains bounds for the pointer's offset from its base object.  Using this information, the verifier
checks various safety properties, such as:

\begin{itemize}[noitemsep]
	\item For every memory access, the operand register $R$ contains a pointer type,
$R \neq \textrm{NULL}$, and $R$ points to within its base object.
	\item If a stack slot is read, then the program has previously written to it.  (This check
prevents leaking of kernel data that was previously stored in that location.)
    \item Pointers are not cast to scalars. (This check prevents kernel addresses from
    leaking to userspace, since such scalars can be stored to a map and read by the user.)
\end{itemize}

eBPF also allows certain program types to access kernel data structures such as socket
and packet objects, whose size may not be known at compile time.  In such cases, the runtime
stores the object size in the program's context, and the verifier checks that any pointer to
these objects undergoes bounds checking before being dereferenced.  We do not discuss this
further, since our attack does not exploit this functionality.

\paragraph{Compile-time Spectre mitigation}
Following the disclosure of Spectre, the eBPF verifier was extended to patch eBPF programs
with run-time mitigations for Spectre bounds check bypass vulnerabilities~\cite{ebpf-spectre-1,ebpf-spectre-2}.
The verifier rewrites any array-map access and any pointer arithmetic operation so that they
are guaranteed to be within the base object's bounds.  For example, \texttt{A[i]} is rewritten
as \texttt{A[i \& (A.size-1)]}, where \texttt{A.size} is rounded up to be a power of 2.  Pointer
arithmetic is handled similarly, leveraging the verifier's knowledge about base objects' size
and pointer offsets.

\subsection{Verifier vulnerability} \label{sec:ebpf-vuln}

When verifying memory safety, the eBPF verifier enumerates only correct execution paths (i.e., that comply with
the semantics of the architecture's instruction set).  The verifier does not reason about the semantically
incorrect execution paths that arise due to (mis)speculative execution.  As a result, the verifier can conclude
that a memory read is safe, but there may be a misspeculated execution flow in which that instruction is unsafe.
\Cref{fig:bpfVulnerableCode} shows an example.

In this example, the semantics of the program are such that the execution of lines~\ref{vulnL3} and~\ref{vulnL5}
is mutually exclusive: line~\ref{vulnL3} executes if and only if \texttt{r0}$=$\texttt{0} and line~\ref{vulnL5} executes
if and only if \texttt{r0}$=$\texttt{1}.  Therefore, the verifier reasons that the only flow in which line~\ref{vulnL5}
executes and memory is read is when \texttt{r6} points to a stack slot, and accepts the program.  Specifically,
when the verifier's analysis reaches line~\ref{vulnL2}, it enumerates two cases:
\begin{itemize}[noitemsep]
\item If the condition on line~\ref{vulnL2} evaluates to TRUE, \texttt{r6}'s type remains a valid stack variable.
\item If the condition on line~\ref{vulnL2} evaluates to FALSE, \texttt{r6}'s type changes to scalar, but the
verifier learns that \texttt{r0} is 0.  Therefore, when it subsequently reaches line~\ref{vulnL4}, it reasons
that the condition there must evaluate to TRUE, and does not consider line~\ref{vulnL5} in these flows.
\end{itemize}

In both cases, every execution flow the verifier considers is safe.  Moreover, the load in line~\ref{vulnL5}
is not rewritten with Spectre mitigation code, because the pointer it dereferences is verified to point to
the stack (and thus within bounds).
Nevertheless, if an attacker
manages to (1) make the dereference of \texttt{r0} (line~\ref{vulnL1}) be a cache miss, so that the
branches take a long time to resolve; and (2) mistrain the branch predictor so that both branches
predict ``not taken'' (i.e., do not execute the \texttt{goto}), then the resulting transient execution sets
\texttt{r6} to the attacker-controlled value in \texttt{r9} (line~\ref{vulnL3}), dereferences that value (line~\ref{vulnL5}),
and leaks it (line~\ref{vulnL6}).

\begin{listing}[t]
\captionsetup[sublisting]{aboveskip=-2pt,belowskip=-3pt}
\begin{sublisting}{.5\textwidth}
\centering
\begin{minted}[mathescape,linenos=false]{c}
// r0 = ptr to a map array entry (verified $\neq$ NULL)
// r6 = ptr to stack slot (verified $\neq$ NULL)
// r9 = scalar value controlled by attacker
\end{minted}
\end{sublisting}\\
\begin{sublisting}{.25\textwidth}
\begin{minted}[linenos]{c}
  r0 = *(u64 *)(r0)|\label{vulnL1}| // miss
A:if r0 != 0x0 goto B|\label{vulnL2}|
  r6 = r9|\label{vulnL3}|
B:if r0 != 0x1 goto D|\label{vulnL4}|
  r9 = *(u8 *)(r6)|\label{vulnL5}|
C:r1 = M[(r9&1)*512];|\label{vulnL6}|//leak
D:...
\end{minted}
\caption{Passes verification.}
\label{fig:bpfVulnerableCode}
\end{sublisting}%
\begin{sublisting}{.25\textwidth}
\begin{minted}{c}
  r0 = *(u64 *)(r0) // miss
A:if r0 == 0x0 goto B
  r6 = r9
B:if r0 != 0x0 goto D
  r9 = *(u8 *)(r6)
C:r1 = M[(r9&1)*512];//leak
D:...
\end{minted}
\caption{Fails verification.}
\label{fig:bpf-non-approved-code}
\end{sublisting}
\caption{eBPF verification vulnerability: leaking a bit (eBPF bytecode; $rX$ stand for eBPF bytecode registers).}
\end{listing}

We remark that in practice, the verifier does not maintain perfect information about each register,
and so may end up enumerating some impossible execution flows.  (I.e., the verifier enumerates an over-approximation
of the correct execution flows.)  Consequently, the verifier inadvertently rejects some speculative type confusion
eBPF gadgets.  \Cref{fig:bpf-non-approved-code} shows an example.  For scalar values, the verifier maintains either
a register's exact value or a possible range of values.  When the verifier considers the case of the condition
on line~\ref{vulnL2} evaluating to FALSE, it cannot track the implication that \texttt{r0}$\neq$\texttt{0}, as that cannot
be represented with a single range.  Since the verifier has no additional information about \texttt{r0}, it goes
on to consider a continuation of the flow in which the condition in line~\ref{vulnL4} also evaluates to FALSE.  In this flow,
line~\ref{vulnL5} dereferences a scalar value, and so the program is rejected.  This rejection is
accidental.  The goal of eBPF developers is to increase the precisions of the verifier's information tracking,
and so under current development trends, we would expect the verifier to eventually accept~\cref{fig:bpf-non-approved-code}.
Improving the verifier's precision is not a problem in and of itself, if eBPF adopts general Spectre mitigations
(i.e., not only bounds enforcing).

\subsection{Proof of concept exploit} \label{sec:ebpf-exploit}

We now describe and evaluate a proof of concept exploit for the eBPF vulnerability.
The exploit implements a universal read gadget~\cite{spectre_google}, which allows
the attacker to read from any kernel virtual address.  This ability allows the
attacker to read all of the machine's physical memory, as the kernel maps all
available physical memory in its virtual address space~\cite{linux-kernel-map}.

The eBPF bytecode is designed to easily map to the x86 architecture, which the eBPF
compiler leverages to map eBPF registers to x86 registers and eBPF branches to
x86 branches.  To guarantee that our victim eBPF program has the required structure
(\cref{fig:bpfVulnerableCode}), we manually encode its bytecode instead of using
a C-to-eBPF compiler.  The kernel then translates the bytecode to native x86 code in
the obvious way.

The main challenge faced by the exploit is how to mistrain the branch predictor, so
that two conditional branches whose ``not taken'' conditions are mutually exclusive
\emph{both} get predicted as ``not taken.''  Two further challenges are (1) how to evict the value
checked by these branches from the cache, so that their resolution is delayed enough
that the misspeculated gadget can read and leak data; and (2) how to
observe the leaked data. These cache-related interactions are non-trivial to perform
because the eBPF program runs in the kernel address space and the attacker, running
in a user process, cannot share memory with the eBPF program. (The main method of
interaction is via eBPF maps, but processes can only manipulate maps using system calls,
not directly.)

\paragraph{Mistraining the branch predictor}
A common branch mistraining technique in Spectre attacks is to repeatedly invoke the
victim with valid input (e.g., an in-bounds array index) to train the branch predictor,
and then perform the attack with an invalid input (e.g., an out-of-bounds array index),
at which point the relevant branch gets mispredicted.  This technique does not apply
in our case: we need to train two branches whose ``not taken'' conditions are mutually
exclusive to both get predicted as ``not taken.''  This means that no matter what input
the eBPF victim gadget is given, at least one of the branches is always taken in its
correct execution.  In other words, we cannot get the branch predictor into a state
that is \emph{inconsistent} with a correct execution by giving it only examples of correct
executions.

To address this problem, we use \emph{cross address-space out-of-place} branch
mistraining~\cite{sok_spectre_meltdown}.  Namely, we set up a ``shadow'' of the
natively-compiled eBPF program in the attacker's process, so that the PHT
entries (\cref{sec:BPU}) used to predict the shadow branches also get used to predict
the victim branches in the kernel.%
\footnote{Note that this approach depends on the branch predictor state not being
cleared when the processor switches to privileged mode on kernel entry.  This is
indeed the case in current processors.}
In our shadow, however, the branches are set up
so the ``not taken'' conditions are not mutually exclusive, and so can be trained
to both predict ``not taken'' (\cref{fig:mistrain}).

\begin{listing}
\renewcommand\fcolorbox[4][]{\textcolor{black}{\strut#4}}
\begin{minted}[texcomments]{c}
// addresses A' and B' collide in the PHT
// with addresses A and B in \cref{fig:bpfVulnerableCode}
A': if r0 == 0x0 goto B'|\label{dummyL2}|
    // dummy register assignment
B': if r0 == 0x0 goto C'|\label{dummyL4}|
    // dummy pointer dereference
C': ...|\label{dummyL5}|
\end{minted}
\vspace{-8pt}
\caption{Mistraining the branch predictor.}
\label{fig:mistrain}
\end{listing}

To ensure PHT entry collision between the shadow and victim branches, we must
control the following factors, upon which PHT indexing is based: (1) the state
of the GHR and (2) the BPU-indexing bits in the branches' virtual addresses.

To control the GHR, we prepend a ``branch slide'' (\cref{fig:branch-slide}) to
both the victim and its shadow. (For the eBPF victim, we generate the branch
slide using appropriate eBPF bytecode, which the kernel compiles into the native
code shown in the listing.) Execution of the branch slide puts the GHR into
the same state both when the shadow is trained and when the victim executes.

To control address-based indexing, we need the address of the shadow gadget
branches to map to the same PHT entries as the victim eBPF program's branches.
We cannot create such a PHT collision directly, by controlling the shadow gadget's
address, since we do not know the victim's address or the hash function used by the branch
predictor. We can, however, perform a ``brute force'' search to find a collision,
as prior work has shown that the PHT on Intel processors has $2^{14}$ entries~\cite{BranchScope}.
We describe our collision search algorithm later, since it relies on our mechanism for leaking data.

\begin{listing}
\renewcommand\fcolorbox[4][]{\textcolor{black}{\strut#4}}
\begin{minted}{gas}
        mov    $0x1,%edi
        cmp    $0x0,%rdi
        jne    |$L_1$|
|$L_1$|:     cmp    $0x0,%rdi
        jne    |$L_2$|
        ...
|$L_{n-1}$|:   cmp    $0x0,%rdi
        jne    |$L_n$|
|$L_n$|:     # exploit starts here
\end{minted}
\vspace{-8pt}
\caption{Branch slide}
\label{fig:branch-slide}
\end{listing}

\paragraph{Cache line flushing}
We need the ability to flush a memory location out of the victim's cache, for two
reasons.  First, we need to cause the read of the value checked by the branches to
miss in the cache, so that the resulting speculation window is large enough to read
and leak the secret data.  Second, one of the simplest ways of leaking the data is
via a {\sc Flush+Reload} cache covert channel~\cite{flush+}, wherein the transient
execution brings a previously flushed secret-dependent line into the cache.  Line~\ref{vulnL6}
in~\cref{fig:bpfVulnerableCode} shows an example, which leaks over an eBPF
array map, \texttt{M}.  Notice that we mask the secret value to obtain a valid offset
into the array, to satisfy the eBPF verifier.  As a result, this example leaks a single bit.

Unfortunately, eBPF programs cannot issue a \texttt{clflush} instruction to flush a
cache line.  This problem can be sidestepped in a number of ways.  We use a clever
technique due to Horn~\cite{jann_ebpf_exploit_poc}.
The basic idea is to perform the required cache line flushes by having \emph{another} eBPF
program, running on a different core, write to these cache lines, which reside in a
shared eBPF array map.  These writes invalidate the relevant cache lines in the victim's cache,
resulting in a cache miss the next time the victim accesses the lines.  After mounting
the attack, the attacker runs a third eBPF program on the victim's core to perform
the timing measurements that deduce which line the transient execution accessed,
and thereby the secret:

\begin{center}
\begin{minipage}{0.44\textwidth}
\renewcommand\fcolorbox[4][]{\textcolor{black}{\strut#4}}
\begin{minted}[texcomments]{c}
r0 = CALL ktime_get_ns()
r1 = M[b]  // b is 0*512 or 1*512
r2 = CALL ktime_get_ns()
return r2 - r0 // if small -> secret is b
\end{minted}
\end{minipage}
\end{center}

This approach leverages the fact that eBPF programs can perform fine-grained timing measurements
by invoking the \texttt{ktime\_get\_ns()} kernel helper.  This is not fundamental for the
attack's success, however.  Similarly to what has been shown for JavaScript~\cite{FantasticTimers},
we could implement a fine-grained ``clock'' by invoking eBPF code on another core to
continuously increment a counter located in a shared eBPF map.

\paragraph{Finding address-based PHT collisions}
To place our shadow branches into addresses that get mapped to the same PHT entries as
the victim's branches (whose address is unknown), we perform the following
search algorithm.

We allocate a 2\,MB buffer and then, for each byte in the buffer, we
copy the shadow gadget to that location and check for a collision by trying the attack.
We first mistrain the branch predictor by repeatedly executing the shadow gadget (whose
branches' PHT entries are hoped to collide with the victim's). We then invoke the in-kernel
victim gadget, configured (by setting the array entry read into \texttt{r0}) so that its correct execution does
not leak (i.e., does not execute line~\ref{vulnL6} in~\cref{fig:bpfVulnerableCode}).
If no leak occurs---i.e., both timing measurements of \texttt{M[b]} indicate a
cache miss---we do not have a collision. If a leak occurs, we may still not have a collision:
the victim may have leaked its own stack variable by executing line~\ref{vulnL5}, either due
to the initial BPU state or if we only have a PHT collision with the second branch.
To rule these possibilities out, we try the attack again, this time with the relevant bit flipped
in that stack variable (which is done by invoking the victim with a different argument).
If the leaked bit flips too, then we do not have a collision; otherwise,
we do.
Once found, a collision can be reused to repeat attacks against the victim.
If the search fails, the attacker can re-load the victim and retry the search.

\subsubsection{Evaluation}

We use a quad-core Intel i7-8650U (Kaby Lake) CPU.
The system runs Ubuntu 18.04.4 LTS with Linux 5.4.11
in a workstation configuration, with applications such as Chrome, TeXstudio, and Spotify running concurrently
to the experiments.

\paragraph{PHT collisions}
We perform 50 experiments, each of which searches for a shadow gadget location that results in PHT collisions with
a freshly loaded victim.
Successful searches take 9.5 minutes on average and occur with 92\% probability (\Cref{table:search-results}).
Our search algorithm can be optimized, e.g., by considering only certain addresses (related to BPU properties and/or kernel buffer alignment).
The search, however, is not a bottleneck for an attack, since once a location for the shadow gadget is found,
it can be reused for multiple leaks. We therefore do not invest in optimizing the search step.

\paragraph{Covert channel quality}
We attempt to leak the contents of one page (4096\,bytes) of kernel memory, which is pre-filled with random bytes.
We leak this page one bit at a time, as described above.  The only difference from~\cref{fig:bpfVulnerableCode}
is that our victim eBPF program receives an argument specifying which bit to leak in the read value, instead of always
leaking the least significant bit.  To leak a bit, we \emph{retry} the attack $k$ times, and output the majority value
leaked over the $k$ retries.

\Cref{table:benchmark-results} shows the resulting accuracy (percentage of bits leaked
correctly) and throughput (bits/second) of the overall attack, as a function of the number of retries.
Since all steps are carried out using the same shadow gadget location, we do not account for the initial search time.
The attack reads from arbitrary memory locations at a rate of 6.7\,KB/sec with 99\% accuracy, and 735\,bytes/sec with 100\% accuracy.
(The success rate with 2 retries is lower due to an implementation artifact in our majority-taking computation.)

\begin{table}
\centering
\small
\begin{tabular}{c c c c c}
\toprule
\textbf{found collision?} & \textbf{average} & \textbf{min.} &  \textbf{max.} & \textbf{median} \\
\midrule
success (46/50) & 9.5\,min. & 20\,sec. & 45\,min. & 8.5\,min. \\
failure (4/50) & \multicolumn{4}{c}{$\approx 53$\,min} \\
\bottomrule
\end{tabular}
\caption{Times to find PHT collision with victim (50 experiments).}
\label{table:search-results}
\end{table}

\begin{table}
\centering
\small
\begin{tabular}{l c c}
\toprule
\textbf{retries} & \textbf{success rate} &  \textbf{transmission rate} \\
\midrule
1 & 99.9\% & 55,416 bps \\
2 & 98.7\% & 28,712 bps \\
10 & 100\% & 5,881 bps \\
100 & 100\% & 584 bps \\
\bottomrule
\end{tabular}
\caption{Accuracy and capacity of the eBPF covert channel.}
\label{table:benchmark-results}
\vspace{-10pt}
\end{table}

\section{Compiler-introduced speculative type confusion} \label{sec:compilers}

In principle, compiler optimizations can create
speculative type confusion gadgets in the emitted code (\cref{lst:spectre-example-tc} shows
a hypothetical example).
Here, we first show that this is not a theoretical concern: deployed compilers can generate speculative type confusion gadgets,
and certain Spectre compiler mitigations do not identify or block such gadgets
(\cref{sec:compiler-poc}).  Motivated by this finding, we perform a binary analysis on
Linux
and find that it contains potential compiler-introduced vulnerabilities
(\cref{sec:compiler-analysis}).

\subsection{Compilers emit gadgets} \label{sec:compiler-poc}

We test different versions of several compilers: GCC, Clang, Intel ICC (from Intel Parallel Studio), and
Microsoft Visual Studio (MSVC).  We find that all of them can compile C code into x86 code that contains
a speculative type confusion gadget.  \Cref{table:compiler-results} summarizes the results.

\Cref{fig:compiler-example} shows an example for GCC; the other compilers produce similar results for similar code.
Here, the code in the first \texttt{if} block overwrites the
\texttt{rdi} register (argument \texttt{p}) with the \texttt{rsi} register (attacker-controlled argument \texttt{x}).
The compiler performs this overwrite because it enables using the same instruction for the assignment to \texttt{foo} at the end of the function. The compiler also reasons that the write to \texttt{*q} might modify \texttt{predicate}
(if \texttt{q} points to \texttt{predicate}), and thus \texttt{predicate} should be re-checked after the first \texttt{if} block.
The compiler's analysis does not understand that in a correct execution, the first \texttt{if} block executing implies that
the second \texttt{if} block does not execute, even if \texttt{q} points to \texttt{predicate}. However, if the attacker mistrains
the branches such that both predict ``not taken,'' the resulting transient execution dereferences the
attacker-controlled value \texttt{x} and leaks its value. Using the mistraining technique of~\cref{sec:ebpf-exploit},
we verify that this is possible.

\paragraph{Spectre mitigations efficacy}
We test whether each compiler's Spectre mitigations apply protection in our example.

\textbf{Clang/LLVM:}
Implements a generic mitigation called speculative load hardening (SLH)~\cite{SLH}.
SLH inserts branchless code that creates a data dependency between each load's address and \emph{all} prior
conditional branch predicates.  SLH thus successfully protects the gadget in our example, but at a high performance
cost (\cref{sec:discussion}).

\textbf{MSVC:} Supports several mitigation levels.  The most aggressive mitigation (\texttt{/Qspectre-load}) inserts an \texttt{lfence} speculation barrier after every load instruction.  This mitigation applies in our example.  However, its documentation warns that
``the performance impact is high''~\cite{spectre-load}.  In contrast, MSVC's recommended Spectre v1 mitigation, \texttt{/Qspectre}~\cite{msvc-spectre-basic},
targets bounds check bypass attacks and does not insert any speculation barriers in our example.

\textbf{ICC:} Similarly to MSVC, ICC supports several mitigation levels~\cite{ICC-mitigation}. It offers two full mitigation options, based on speculation barriers (\texttt{all-fix}) or SLH (\texttt{all-fix-cmov}), with the former documented as having ``the most run-time performance cost''~\cite{ICC-mitigation}. Both options apply in our example. ICC also offers a ``vulnerable code pattern'' mitigation, which does not insert speculation barriers in our example.

\textbf{GCC:} Does not support a whole-program mitigation. It offers a compiler intrinsic
for safely accessing values in the face of possible misspeculation.  However, programmer who equate Spectre v1 with
bounds check bypass have no reason to use this intrinsic in our example, so we consider GCC's mitigation inapplicable
in our case.

\begin{table}
\centering
\small
\begin{tabular}{p{.15\textwidth} c c}
\toprule
\textbf{compiler} & \textbf{emits gadget?} &  \textbf{mitigates gadget?} \\
\midrule
Clang/LLVM (v3.5, v6, v7.01, v10.0.1) & yes & yes \\
MSVC (v16) & yes & suggested: no; full: yes \\
ICC (v19.1.1.217) & yes & lightweight: no; full: yes \\
GCC (v4.8.2, v7.5.0) & yes & N/A \\
\bottomrule
\end{tabular}
\caption{Compilers introducing speculative type confusion.}
\label{table:compiler-results}
\end{table}

\begin{listing}[t]
\renewcommand\fcolorbox[4][]{\textcolor{black}{\strut#4}}
\captionsetup[sublisting]{aboveskip=-2pt}
\begin{sublisting}{.24\textwidth}
\begin{minted}{c}
volatile char A[256*512];
bool predicate;
char* foo;

void victim(char *p,
            uint64_t |\colorbox{attackRed}{\textbf{x}}|,
            char *q) {
  unsigned char v;

  if (predicate) {
    p = (char *) |\colorbox{attackRed}{\textbf{x}}|;
    *q |= 1;
  }
  if (!predicate) {
    v = A[(*p) * 512];
  }
  foo = p;
}
\end{minted}
\caption{C code}
\end{sublisting}%
\begin{sublisting}{.26\textwidth}
\begin{minted}[texcomments]{gas}
# args: p in %rdi
#       \colorbox{attackRed}{\textcolor{black}{\textbf{x}}} in \colorbox{attackRed}{\textcolor{black}{\textbf{%rsi}}}
#       q in %rdx

   # first "if":
   cmpb   $0x0,(predicate)
B1:je     L1 # skip 2nd if
   # assignment to p:
   mov    |\colorbox{attackRed}{\textcolor{black}{\textbf{%rsi}}}|,%rdi
   # assignment to q:
   orb    $0x1,(%rdx)
   # second "if":
   cmpb   $0x0,(predicate)
B2:jne    L2
   # deref p & leak
L1:movsbl (%rdi),%eax
   shl    $0x9,%eax
   cltq
   movzbl A(%rax),%eax
L2:mov    %rdi,(foo)
   retq
\end{minted}
\caption{Emitted x86 code.}
\end{sublisting}
\caption{Example of C code compiled into a speculative type confusion gadget (GCC 4.8.2, -O1).
Argument \colorbox{attackRed}{\textbf{x}} is attacker-controlled.}
\label{fig:compiler-example}
\end{listing}

\subsection{Finding compiler-introduced gadgets} \label{sec:compiler-analysis}

To find potential compiler-introduced speculative type confusion vulnerabilities in the wild,
we perform a binary-level static analysis of Linux 5.4.11, compiled with different GCC versions and optimization flags.

\paragraph{Goal \& methodology}
We set to find out if the kernel can be maneuvered (via transient execution) to dereference a user-supplied address,
which is the core of the attack.  We explicitly do not consider if or how the result of the dereference can be leaked,
for the following reasons.  Once a secret enters the pipeline, it can be leaked in many ways, not necessarily over
a cache covert channel (e.g., port contention~\cite{smother} or execution unit timings~\cite{netspectre}). It is
beyond our scope to exhaustively evaluate all possible leaks to determine if a ``confused'' dereference can be
exploited.  Also, dereferences that appear unexploitable on our test setup may be exploitable with a different combination
of kernel, compiler, and flags.  Finally, today's unexploitable dereferences may become exploitable in the future, due to
(1) discovery of new microarchitectural covert channels, (2) secrets reaching more instructions on future processors with deeper
speculation windows, or (3) kernel code changes.  Overall, the point is: \emph{the architectural contract gets breached
when the kernel dereferences a user-supplied address}.  We thus focus on detecting these breaches.

\paragraph{Analysis}
We use the Radare2 framework~\cite{radare2} for static analysis and Triton~\cite{SSTIC2015-Saudel-Salwan} to perform
taint tracking and symbolic execution.
Conceptually, for each system call, we explore all possible execution paths that start at the system call's
entry point and end with its return.  We look for loads whose operand is unsafe (user-controlled) in one execution but
safe (kernel-controlled) in another execution.  We detect such loads by executing each path while performing taint
analysis.  We maintain a taint bit for each architectural register and each memory word.  When analyzing a system
call, we maintain two sets: $U$ and $K$, initially empty.  For each path, we initially taint the system call's arguments
(which are user-controlled) and then symbolically execute the path while propagating taint as described below.
When we encounter a load instruction, we place it into $U$ if its operand is tainted, or into $K$ otherwise.
We flag every load $L \in U \cap K$. \Cref{fig:registers-reuse-analysis-code} shows pseudo code of the algorithm.

Our analysis is designed for finding proofs of concept.  As explained next, the analysis is not complete (may miss potential vulnerabilities),
and it is not sound (may flag a load that is not a potential vulnerability).  The results we report are after manually discarding such
false positives.

\textbf{Incomplete:} We often cannot explore all possible execution flows of a system call, since their number is
exponential in the number of basic blocks.  We therefore skip some paths, which means we may miss
potential vulnerabilities.  We limit the number of paths analyzed in each system call in two ways.

First, we limit the number of explored paths but ensure that every basic block is covered. We start with the set $Paths$
of the 1000 longest acyclic paths, to which we add the longest path $P \not\in Paths$ that contains the basic block $B$, for
each basic block $B$ not already covered by some path in $Paths$. The motivation for adding these latter paths is to
not ignore possible flows to \emph{other} basic blocks; loads in a basic block covered by a single explored
path cannot themselves be identified as vulnerable.

Second, when exploring paths, we do not descend into called functions (i.e., skip \texttt{call}
instructions).  Instead, we queue that called function and the taint state, and analyze them independently later.
Overall, \emph{our analysis not finding potential vulnerabilities does not imply that none exist.}

\textbf{Unsound:} The analysis is unsound because it abstracts memory and over-approximates taint.
We model kernel and user memory as unbounded arrays, $M^K$ and $M^U$, respectively.
Let $T(x)$ denote the taint of $x$, where $x$ is either a register or a memory location.
Values in $M^U$ are always tainted ($\forall a: T(M^U[a])=1$), whereas values in $M^K$ are initially untainted, but may
become tainted due to taint propagation.  We execute a memory access with an untainted operand on $M^K$, and from $M^U$ otherwise.
A store $M^K[a] = R$ sets $T(M^K[a])=T(R)$.  A load $R=M^K[a]$ sets $T(R)=T(M^K[a])$, and similarly for loads from $M^U$.
We assume that reads of uninitialized memory locations read 0. As a result, many objects in the analyzed execution
appear to alias (i.e., occupy the same memory locations), which can lead to inaccurate taint propagation.
For instance, in the following code, a tainted value is stored into $M^K[0x10]$ and taints it,
which causes the result of the subsequent load that reads $M^K[0x10]$ to be tainted. In the real execution, however,
the store and load do not alias as they access different objects.
\begin{center}
\begin{minipage}{.4\textwidth}
\renewcommand\fcolorbox[4][]{\textcolor{black}{\strut#4}}
\begin{minted}{gas}
   mov    |\colorbox{attackRed}{\textcolor{black}{\textbf{%rax}}}|,0x10(%rbx)   # p->foo = x
   mov    0x10(%rcx),%rdx     # v = q->bar
\end{minted}
\end{minipage}
\end{center}

Due to its unsoundness, we manually verify every potential speculative type confusion flagged by the analysis.
Because we limit the number of explored paths, the number of reports (and thus false positives) is not prohibitive
to inspect.

\begin{listing}
\begin{minted}{python}
analyze_syscall(S):
  U = K = |$\emptyset$|
  G = control-flow graph of S
  for each acyclic path P |$\in$| G:
     analyze_path(P)

analyze_path(P):
  reset state, taint input argument registers
  for each instruction I in P:
    # propagate taint
    if I is a load/store: propagate taint from/to memory
    else: taint the output register of I iff
          one of its operands is tainted
    symbolically execute I
    if I is a load:
      add I to U or K, as appropriate
      flag I if I |$\in$| U |$\cap$| K
\end{minted}
\vspace{-20pt}
\caption{Finding potential compiler-introduced speculative type confusion. (Conceptual algorithm, see text for optimizations.)}
\label{fig:registers-reuse-analysis-code}.
\end{listing}

\subsection{Analysis results} \label{sec:analysis-results}

We analyze Linux v5.4.11, compiled with GCC 5.8.2 and 9.3.0. We use the \texttt{allyes} kernel configuration, which enables every
configuration option, except that we disable the kernel address sanitizer and stack leak prevention, as they instrument code
and so increase the number of paths to explore. The case studies below are valid with these options enabled.  We build the kernel
with the \texttt{-Os} and \texttt{-O3} optimization levels.  We analyze every system calls with arguments (393 in total). \Cref{table:binary-analysis-results}
summarizes the results.

Depending on the optimization level, GCC 9.3.0 introduces potential gadgets into 2--20 system calls, all of which
stem from the same optimization (\cref{sec:gcc930}). GCC 5.8.2 introduces a ``near miss'' gadget into one system call
(\cref{sec:keyctl}). This gadget is not exploitable in the kernel we analyze, but the pattern exhibited would be
exploitable in other cases.

All the gadgets found are traditionally considered not exploitable,
as the mispredicted branches depend on registers whose value is available (not cache misses), and so can resolve quickly.
We show, however, that branches with available predicates \emph{are} exploitable on certain x86 processors (\cref{sec:amd}).

\begin{table}
\centering
\small
\begin{tabular}{c c c}
\toprule
\textbf{compiler} & \textbf{flags} & \textbf{\# vulnerable syscalls} \\
\midrule
GCC 9.3.0 & -Os & 20 \\
GCC 9.3.0 & -O3 & 2 \\
GCC 5.8.2 & -Os & 0$^\dagger$ \\
GCC 5.8.2 & -O3 & 0 \\
\bottomrule
\multicolumn{3}{p{.35\textwidth}}{$^\dagger$ One potential vulnerability exists,
see~\cref{sec:keyctl}.}
\end{tabular}\\
\caption{Compiler-introduced speculative type confusion in Linux.}
\label{table:binary-analysis-results}
\vspace{-8pt}
\end{table}

\subsubsection{Supposedly NULL pointer dereference} \label{sec:gcc930}

The gadgets introduced by GCC 9.3.0 all stem from the same pattern, a simplified example of which appears in~\cref{fig:registers-ruse}.
The system call receives an untrusted user pointer, \texttt{uptr}.  If \texttt{uptr} is not NULL, it safely copies its contents into
a local variable. Next, the system call invokes a helper \texttt{f}, which receives NULL if \texttt{uptr} was NULL, or a pointer to
the kernel's local variable otherwise.  The helper \texttt{f} (not shown) thus expects to receive a (possibly-NULL) kernel pointer,
and therefore dereferences it (after checking it is non-NULL).

In the emitted code, the compiler introduces a speculative type confusion gadget by reusing \texttt{uptr}'s register to
pass NULL to \texttt{f} when \texttt{uptr} is NULL. If the attacker invokes the system call with a non-NULL \texttt{uptr} and
the NULL-checking branch mispredicts ``taken'' (i.e., \texttt{uptr}$=$NULL), then \texttt{f} gets called with the
attacker-controlled value and dereferences it.

It is not clear that the gadget can be exploited, as both the mispredicted branch and the dereference depend on
the same register.  Why would the processor execute the dereference if it knows that the branch mispredicted?  We discuss
this in~\cref{sec:amd}.

\begin{listing}[t]
\renewcommand\fcolorbox[4][]{\textcolor{black}{\strut#4}}
\captionsetup[sublisting]{aboveskip=-2pt}
\begin{sublisting}{.23\textwidth}
\begin{minted}{c}
syscall(foo_t* |\colorbox{attackRed}{\textbf{uptr}}|) {
  foo_t kfoo;
  // some code
  if (|\colorbox{attackRed}{\textbf{uptr}}|)
    copy_from_user(&kfoo,
                   |\colorbox{attackRed}{\textbf{uptr}}|,
                   ...);
  f(|\colorbox{attackRed}{\textbf{uptr}}| ? &kfoo : NULL);
  // rest of code
}
\end{minted}
\caption{C pattern}
\end{sublisting}%
\begin{sublisting}{.27\textwidth}
\begin{minted}[texcomments]{gas}
# args: uptr in \colorbox{attackRed}{\textcolor{black}{\textbf{\%rdi}}}
  ...
  testq |\colorbox{attackRed}{\textcolor{black}{\textbf{\%rdi}}}||,\colorbox{attackRed}{\textcolor{black}{\textbf{\%rdi}}}|
  je L # jump if \colorbox{attackRed}{\textcolor{black}{\textbf{\%rdi}}}==0
  # set copy\_from\_user args
  ...
  # \%rbp contains addr of
  # stack buffer
  mov %rbp, %rdi
  call copy_from_user
L:callq f
\end{minted}
\caption{Emitted x86 code.}
\end{sublisting}
\caption{Reusing registers for a function call.}
\label{fig:registers-ruse}
\end{listing}

\subsubsection{Stack slot reuse} \label{sec:keyctl}

GCC 5.8.2 with the \texttt{-Os} (optimize for space) flag introduces an interesting gadget.
The gadget instance we find is ``almost'' exploitable.  We describe it not only
to show how a small code change could render the gadget exploitable, but also to demonstrate how
insidious a compiler-introduced gadget can be, and how difficult it is for programmers to
reason about.

The gadget (\cref{fig:registers-ruse-outparam}) is emitted into a function called by the \texttt{keyctl} system call.
In this function, the compiler chooses to allocate space for the stack slot of a local variable
(\texttt{dest\_keyring}) by pushing the \texttt{rcx} register to the stack (a one-byte opcode)
instead of subtracting from the stack pointer (a four-byte opcode).  The \texttt{rcx} register holds a
user-controlled value, one of the caller's (\texttt{keyctl}) arguments.  The code then calls
\texttt{get\_instantiation\_keyring()}, passing it the address of \texttt{dest\_keyring}.
If \texttt{get\_instantiation\_keyring()} returns successfully, the code calls
\texttt{key\_instantiate\_and\_link()}, which dereferences \texttt{dest\_keyring}.

\begin{listing}
\renewcommand\fcolorbox[4][]{\textcolor{black}{\strut#4}}
\captionsetup[sublisting]{aboveskip=-2pt}
\begin{sublisting}{.5\textwidth}
\begin{minted}{c}
long keyctl_instantiate_key_common(key_serial_t id,
				   struct iov_iter *from,
				   key_serial_t ringid) {
 struct key *dest_keyring;
 // ... code ...
 ret = get_instantiation_keyring(ringid,rka,&dest_keyring);
 if (ret < 0)
  goto error2;
 ret = key_instantiate_and_link(rka->target_key, payload,
                                plen, dest_keyring,
                                instkey);
 // above call dereferences dest_keyring
}
\end{minted}
\caption{C code}
\end{sublisting}\\
\begin{sublisting}{.5\textwidth}
\begin{minted}[texcomments]{gas}
  # \colorbox{attackRed}{\textcolor{black}{\textbf{\%rcx}}} is a live register from caller
  push |\colorbox{attackRed}{\textcolor{black}{\textbf{\%rcx}}}|
  # ... code ...
  lea    0x18(%r14),%rsi  # rka argument
  mov    %rsp,%rdx        # \&dest\_keyring argument
  mov    %r15d,%edi       # ringid argument
  callq  get_instantiation_keyring  # returns error
  test   %rax,%rax        # if (ret < 0)
  mov    %rax,%rbx
  js     error2           # mispredict no error
  ...
  mov    (%rsp),%rcx      # dest\_keyring argument
  # dest\_keyring could be old \colorbox{attackRed}{\textcolor{black}{\textbf{\%rcx}}} if not
  # overwritten in get\_instantiation\_keyring()
  callq  key_instantiate_and_link
\end{minted}
\caption{Emitted x86 code.}
\end{sublisting}
\caption{Attacker-controlled stack slot reuse in the \texttt{keyctl} system call flow.}
\label{fig:registers-ruse-outparam}
\end{listing}

In order to pass \texttt{dest\_keyring} to \texttt{key\_instantiate\_and\_link()}, the code reads
its value from the stack.  Consider what happens if the earlier \texttt{get\_instantiation\_keyring()} call
encounters an error, and therefore leaves the stack slot with its original (user-controlled)
value. In a correct execution, \texttt{key\_instantiate\_and\_link()} never gets called,
due to the error-checking flow. But if an attacker induces the error-checking branch to mispredict,
\texttt{key\_instantiate\_and\_link()} gets called with a user-controlled pointer
argument to dereference.  The only reason this instance is not exploitable is that
\texttt{get\_instantiation\_keyring()} error flows overwrite \texttt{dest\_keyring}.
There is no security-related reason for this overwrite, since \texttt{dest\_keyring} is a local variable
that is never exposed to the user (i.e., there is no potential kernel information leak).
Were this function to use a different coding discipline, the gadget would be potentially exploitable.

\subsection{Potential exploitability of the gadgets} \label{sec:amd}

Exploiting the gadgets described in~\cref{sec:analysis-results} appears impossible.
Typical Spectre gadgets exploit branches whose condition depends on values being fetched
from memory, and so take a long time to resolve.  In our case, the branch \emph{still
mispredicts}, since prediction is done at fetch time (\cref{sec:BPU}).  However,
the branch condition is computable immediately when it enters the processor backend,
as the values of the registers involved are known.  One would expect the processor to
immediately squash all instructions following the branch once it realizes that the
branch is mispredicted, denying any subsequent leaking instructions (if they exist) a chance to execute.

The above thought process implicitly assumes that the processor squashes instructions
in the shadow of a mispredicted branch when the branch is \emph{executed}.  But what if
the squash happens only when the branch is \emph{retired}?  A branch's retirement can
get delayed if it is preceded by a long latency instruction (e.g., a cache missing load), which
would afford subsequent transient instructions a chance to execute.

We test the above hypothesis and find it to be false, but in the process, we discover
that some x86 processors exhibit conceptually similar behavior due to other reasons.%
\footnote{We did not test non-x86 processors.}
Namely, we find that how x86 processors perform a branch misprediction squash depends
in some complex way on \emph{preceding branches} in the pipeline.  Specifically,
the squash performed by a mispredicting branch $B_1$ can get delayed if there is a
preceding branch $B_0$ whose condition has not yet resolved.

\Cref{fig:amd-example} shows the experiment.  We test a gadget similar to the ``supposedly
NULL dereference'' (\cref{sec:gcc930}) gadget.  We train the victim so that both branches
are taken (\texttt{*p==1, m!=bad\_m}).  We then invoke it so that both mispredict
(\texttt{*p==0, m==bad\_m}), with \texttt{p} being flushed from the cache, and test
whether \texttt{m} is dereferenced and its value \texttt{s} is leaked.  \Cref{table:amd-results}
shows the results: a leak can occur on both Intel and AMD processors, but its probability
is minuscule on Intel processors. The small success probability and its dependence on the
exact instructions in the gadget indicate that the leak occurs due some complex microarchitectural
interaction.

\paragraph{Implications} The fact that leaks can be realistically observed (for perspective: on AMD processors,
our experiment observes $\approx 10$\,K leaks per minute) means that compiler-introduced gadgets are a
real risk.  For any gadget instance, the kernel's flow may be such that there are slow-to-resolve
branches preceding the gadget, and/or the attacker may be able to slow down resolution of preceding
branches by evicting data from the cache.

\begin{listing}[t]
\renewcommand\fcolorbox[4][]{\textcolor{black}{\strut#4}}
\captionsetup[sublisting]{aboveskip=-2pt}
\begin{sublisting}{.2\textwidth}
\begin{minted}{c}
int victim(int* p,
           T *m,
           T *bad_m,
           char *A) {
 if (*p == 1) {
  if (m != bad_m) {
   T s = *m;
   A[s*0x1000];
  }
  return 5;
 }
 return 0;
}
\end{minted}
\caption{C code}
\end{sublisting}%
\begin{sublisting}{.3\textwidth}
\begin{minted}[texcomments]{gas}
   # deref *p (cache miss)
   mov   (%rdi),%edi
   mov   $0x0,%eax
   cmp   $0x1,%edi   # *p==1 ?
   je    L2   # jmp if *p==1
L1:repz retq
L2:mov  $0x5,%eax
   cmp   %rdx,%rsi # m==bad\_m ?
   je    L1   # jmp if m==bad\_m
   movzbl (%rsi),%eax # s = *m
   shl   $0xc,%eax
   cltq
   add   %rax,%rcx
   movzbl (%rcx),%eax # leak s
   mov   $0x5,%eax
   jmp   L1
\end{minted}
\caption{Emitted x86 code (\texttt{T}=\texttt{char}).}
\end{sublisting}
\caption{Evaluating processor branch misprediction squashes.}
\label{fig:amd-example}
\end{listing}

\begin{table}
\centering
\resizebox{\columnwidth}{!}{%
\begin{tabular}{l c c}
\toprule
\multicolumn{1}{c}{\multirow{1}{*}{\textbf{processor}}} & \multicolumn{2}{c}{\textbf{leak probability}} \\
                                    & \textbf{\texttt{T=char}} & \textbf{\texttt{T=long}} \\
\midrule
AMD EPYC 7R32                       & $1/10^5$ & $1/5000$ \\
AMD Opteron 6376                    & $1/10^5$ & $1/5000$ \\
Intel Xeon Gold 6132 (Skylake)      & $1/(5.09 \times 10^7$) &  $1/(1.36 \times 10^6)$    \\
Intel Xeon E5-4699 v4 (Broadwell)   &  $1/(3.64 \times 10^9)$    & $1/(6.2 \times 10^9)$    \\
Intel Xeon E7-4870 (Westmere)       &  $1/(1.67 \times 10^9)$    & $1/(2.75 \times 10^7)$    \\
\bottomrule
\end{tabular}%
}
\caption{x86 branch squash behavior (in 30\,B trials).}
\label{table:amd-results}
\vspace{-12pt}
\end{table}

\section{Speculative polymorphic type confusion} \label{sec:jumpswitch}

Linux defends from indirect branch target misspeculation attacks (Spectre-BTB)
using \emph{retpolines}~\cite{retpoline}.  A retpoline replaces an indirect branch
with a thunk that jumps to the correct destination in a speculation-safe way,
but incurs a significant slowdown~\cite{234862}.
Since the original Spectre-BTB attacks diverted execution to \emph{arbitrary} locations,
retpolines can appear as an overly aggressive mitigation, as they block all branch
target speculation instead of restricting it to \emph{legal} targets~\cite{234862}.

Accordingly, Linux is moving in the direction of replacing certain retpoline call sites
with thunks of conditional \emph{direct} calls to the call site's most likely targets,
plus a retpoline fallback~\cite{retpoline-pain}.  \Cref{fig:jumpswitch-thunk}
shows an example.  JumpSwitches~\cite{234862} take the idea further, and propose to
dynamically promote indirect call sites into such thunks by learning probable targets
and patching the call site online.

In this section, we detail how this approach can
create speculative type confusion vulnerabilities (\cref{sec:jumpswitch-vuln}) and analyze the prevalence
of such issues in Linux (\cref{sec:smatch})

\begin{listing}
\renewcommand\fcolorbox[4][]{\textcolor{black}{\strut#4}}
\begin{minted}{gas}
# %rax = branch target
  cmp $0xXXXXXXXX, %rax  # target1?
  jz $0xXXXXXXXX
  cmp $0xYYYYYYYY, %rax  # target2?
  jz $0xYYYYYYYY
  ...
  jmp ${fallback} # jmp to retpoline thunk
\end{minted}
\caption{Conditional direct branches instead of indirect branch.}
\label{fig:jumpswitch-thunk}
\vspace{-8pt}
\end{listing}

\subsection{Virtual method speculative type confusion} \label{sec:jumpswitch-vuln}

It has been observed (in passing) that misprediction of an indirect call's target can lead
to speculative type confusion in object-oriented polymorphic code~\cite{sok_spectre_meltdown}.
The problem occurs when a branch's valid but wrong target, $f$, is speculatively invoked
instead of the correct target, $g$.  The reason that both $f$ and $g$ are valid targets is
that both implement some common interface.  Each function, however, expects some or all of
its arguments to be a different \emph{subtype} of the types defined by the interface.  As
a result, when the misprediction causes $f$'s code to run with $g$'s arguments, $f$ might
\emph{derive} a variable $v$ of type $T_f$ from one of $g$'s arguments, which is really of
type $T_g$.

Prior work~\cite{MSFT_stc,javascript-stc} describes how, if $T_g$ is smaller than $T_f$,
$f$ might now perform an out-of-bounds access.  We observe, however, that the problem is more
general.  Even if both types are of the same size, $f$ might still dereference a field in $T_f$
which now actually contains some user-controlled content from a field in $T_g$, and subsequently
inadvertently leak the read value.  Moreover, the problem is transitive: $f$ might dereference a field
that is a pointer in both $T_f$ and $T_g$, but points to an object of a different type in each,
and so on.

In the following sections, we analyze the prevalence of potential polymorphism-related speculative
type confusion is Linux.  Our analysis is forward looking: we explore \emph{all} indirect call sites, not
only the ones that have already been converted to conditional call-based thunks.  Such a
broad analysis helps answering questions such as: How likely is it that manually converting a
call site (the current Linux approach) will create a vulnerability? What are the security
implications of adopting an ``all-in'' scheme likes JumpSwitches?

\subsection{Linux analysis} \label{sec:smatch}

Linux makes heavy use of polymorphism and data inheritance (subtype derivation) in implementing
internal kernel interfaces.  (Linux code implements inheritance manually, due to being in C,
as explained below.)  We perform a source-level vulnerability analysis by extending the \texttt{smatch} static
analysis tool~\cite{smatch}.  As in~\cref{sec:compilers}, we do not claim that if our analysis
finds no vulnerabilities then none exist.

At a high-level, the analysis consists of four steps, detailed below.
(\Cref{fig:smatch-anal-pseudo-code} shows pseudo code.)

\begin{listing}
\begin{minted}{python}
# $\texttt{\textbf{targets}}$: a mapping from function pointer
# fields in types to their valid targets
# $\texttt{\textbf{derivedObjs}}$: a mapping from function
# arguments to possible private structs they
# derive
 # $\circled{1}$ find call site target
 for every assignment |$x.a = g$| where |$g$| is a function
     and |$x$| is of type |$T$|
  targets[|$T,a$|].add(|$g$|)
 # $\circled{2}$ find derived_objects
 for every |$g$| in targets, scan control-flow graph of |$g$|:
  if |$i$|-th arg of |$g$| is used to derive struct of type |$T$|:
   derivedObjs[|$g, i$|] = |$T$|
 # $\circled{3}$ find all overlaps
 overlaps = set()
 for every |$T,a \mapsto \{g_1,\dots,g_m\}$| in targets:
  for every pair (|$g_i, g_j$|):
   for every |$g_i,a \mapsto D_i$| in derivedObjs:
    for every field |$f_i$| of |$D_i$| that is user-controllable:
     |$D_j$| = derivedObjs[|$g_j,a$|]
     let |$f_j$| be the overlapping field in |$D_j$|
     overlaps.add(|$(g_i,D_i,f_i,g_j,D_j,f_j)$|]
 # $\circled{4}$ find potentially exploitable overlaps
 for each |$(g_i,D_i,f_i,g_j,D_j,f_j)$| in overlaps:
  scan control-flow graph of |$g_j$|
  if |$D_j.f_j$| is dereferenced:
   let v be the data read from |$D_j.f_j$|
   if v is used to index an array or v is dereferenced:
     |\texttt{\textbf{flag}}| |$\bm{(g_i,D_i,f_i,g_j,D_j,f_j)}$|
\end{minted}
\caption{Finding potential speculative polymorph type confusion.}
\label{fig:smatch-anal-pseudo-code}
\end{listing}

\noindent\textbf{\circled{1} Find legal targets:}
For each structure type and each function pointer field in that type, we build a set of legal targets
that this field might point to.

\noindent\textbf{\circled{2} Identify subtype derivations:}
For each function $g$ that is a legal target of some call site, we attempt
to identify the arguments used to derive $g$-specific (subtype) objects.  Since Linux implements data
inheritance manually, we scan for the relevant patterns (illustrated in~\cref{fig:inheritance-patterns}):
(1) a ``private data'' field in the parent structure points to the derived
object (\cref{fig:private-data}); (2) the derived object is the first field in the parent,
and obtained by casting (\cref{fig:casting}); and (3) the derived object is some field in the
parent, extracted using the \texttt{container\_of} macro (\cref{fig:containted-struct}).

\begin{listing}
\captionsetup[sublisting]{aboveskip=-5pt}
\begin{sublisting}{.25\textwidth}
\begin{minted}{c}
struct Common {
 void* private;
};
struct Derived {...};

void foo(Common* c) {
 Derived* d = c->private;
\end{minted}
\caption{Private field.}
\label{fig:private-data}
\end{sublisting}%
\begin{sublisting}{.25\textwidth}
\begin{minted}{c}
struct Common {...}
struct Derived {
 struct Common common_data;
 ...
}
void foo(Common* c) {
 Derived* d = (Derived*) c;
\end{minted}
\caption{Casting.}
\label{fig:casting}
\end{sublisting}\\
\begin{sublisting}{.5\textwidth}
\begin{minted}{c}
struct Derived {
 ...
 struct Common common;
 ...
}
void foo(Common* c) {
 Derived* d = container_of(c, Derived*, common);
\end{minted}
\caption{Contained structs.}
\label{fig:containted-struct}
\end{sublisting}%
\caption{Linux data inheritance patterns.}
\label{fig:inheritance-patterns}
\end{listing}

\noindent\textbf{\circled{3} Find overlapping fields:}
This is a key step.  For every pair of functions that are legal targets of some call site,
we search for \emph{overlapping} fields among the objects derived from the same function argument.
Two fields overlap if (1) their (start,end) offset range in the respective object types intersect,
(2) one field is user-controllable, and (3) the other field, referred to as the
\emph{target}, is not user-controllable.  We rely on \texttt{smatch} to identify which
fields are user-controllable, which is done based on Linux's \texttt{\_\_user} annotation~\cite{user-attrib}
and heuristics for tracking untrusted data, such as network packets.  An overlap where
the target field is a kernel pointer can potentially lead to an attacker-controlled dereference.

\noindent\textbf{\circled{4} Search for vulnerabilities:}
This steps takes a pair of functions $g_i,g_j$ identified as using derived objects with overlapping
fields, and tries to find if the overlaps are exploitable.  We run a control- and data-flow
analysis on $g_j$, the function using the object with the target field, and check
if that field is dereferenced.
This process finds thousands of potential attacker-controlled
dereferences.  To make manual verification of the results tractable, we try to detect if the
value read by the dereference gets leaked via a cache covert channel.  We consider two
types of leaks: if some array index depends on the value, and a ``double dereference'' pattern, in
which the value is a pointer that is itself dereferenced. The latter pattern can
be used to leak the L1-indexing bits of the value.

\subsection{Analysis results} \label{sec:smatch-results}

We analyze Linux v5.0-rc8 (default configuration) and v5.4.11 (\texttt{allyes} configuration).
\Cref{table:jumpswith-stats} summarizes the results.  While we find thousands of potential
attacker-controlled dereferences, most are double dereferences, which we do not
consider further.
Manual inspection of the array indexing cases reveals that they are \emph{latent} vulnerabilities,
which are not (or likely not) exploitable, but could become exploitable by accident:
\begin{itemize}[noitemsep,leftmargin=*]
\item Most cases let the attacker control $<64$ bits of the target pointer, with which it cannot
represent a kernel pointer (e.g., attacker controls a 64-bit field in its structure, but it only
overlaps the target field over one byte).  A change in structure layout or field size could make these
cases exploitable.
\item In other cases, the attacker does not have full control over its field (e.g., it is a
flag set by userspace that can only take on a limited set of values).
\Cref{sec:case-study} shows an example of such a case.
A change in the semantics of the field could render these cases exploitable.
\item Some cases are false positives due to imprecision of the analysis (e.g., a value read and
used as an array index is masked, losing most of the information).
\item Finally, we discard some cases because there is a function call between the dereference
and the array access, so we assume that the processor's speculation window in insufficient
for the value to reach the array access.
\end{itemize}

Note that we may be over-conservative in rejecting cases. The reason that we do not invest
in exploring each case in detail is because we are looking at an analysis of all indirect calls,
most of which are currently protected with retpolines.  The takeaway here is that were a
conditional branch-based mitigation used instead of retpolines, the kernel's security would be on
shaky ground.

\begin{table}
\centering
\small
\begin{tabular}{l r r}
\toprule
                     & \multicolumn{1}{c}{\textbf{5.0-rc8 (def.)}} & \multicolumn{1}{c}{\textbf{5.4.11 (allyes)}} \\
\midrule
\textbf{\# flagged}             & 2706              & 8814  \\
\textbf{double deref}           & 2578              & 8512 \\
\textbf{array indexing}         & 128               & 302 \\
\midrule
\textbf{array indexing exploitable?}      &                   & \\
\textbf{no: $\bm{<64}$ bit overlap} & 108               & 261 \\
\textbf{no: limited control}        &  11               &  13 \\
\textbf{no: other}                  &   4               &  17 \\
\textbf{no(?): speculation window}    &   5               &  11 \\
\bottomrule
\end{tabular}
\caption{Potential speculative polymorph data confusion in Linux.}
\label{table:jumpswith-stats}
\vspace{-12pt}
\end{table}

\subsection{Case study of potential vulnerability} \label{sec:case-study}

\begin{listing}[t]
\captionsetup[sublisting]{aboveskip=-5pt}
\begin{sublisting}{.5\textwidth}
\begin{minted}{c}
ssize_t port_type_show(struct device *dev,
                       struct device_attribute *attr,
                       char *buf) {
 // container_of use
 struct typec_port *port = to_typec_port(dev);
 if (port->|\colorbox{attackRed}{\textbf{cap}}|->type == TYPEC_PORT_DRP)
  return ...;
 return sprintf(buf, "[%s]\n",
          typec_port_power_roles[port->|\colorbox{attackRed}{\textbf{cap}}|->type]);
}
\end{minted}
\caption{Mispredicted target.}
\label{fig:analysis-example-target}
\end{sublisting}\\
\begin{sublisting}{.5\textwidth}
\begin{minted}{c}
ssize_t max_freq_show(struct device *dev,
                      struct device_attribute *attr,
                      char *buf) {
 // container_of use
 struct devfreq *df = to_devfreq(dev);
 return sprintf(buf, "%lu\n", min(df->scaling_max_freq,
                              df->max_freq));
}
\end{minted}
\caption{Actual target.}
\label{fig:analysis-example-actual}
\end{sublisting}\\
\begin{sublisting}{.5\textwidth}
\begin{minted}{c}
ssize_t max_freq_store(struct device *dev,
                       struct device_attribute *attr,
                       const char *buf, size_t count) {
 ...
 if (freq_table[0] < freq_table[df->profile->max_state - 1])
  value = freq_table[df->profile->max_state - 1];
 else
  value = freq_table[0];
  ... value is stored into max_freq ...
}
\end{minted}
\caption{Value of \texttt{max\_freq} is constrained.}
\label{fig:analysis-example-write}
\end{sublisting}
\caption{Speculative polymorphic type confusion case study.}
\label{fig:analysis-example}
\end{listing}

To get a taste for the difficulty of reasoning about this type of speculative type confusion,
consider the example in~\cref{fig:analysis-example}.  The functions in question belong the USB
type C driver and to the \texttt{devfreq} driver.  They implement the \texttt{show} method of
the driver's attributes, which is used to display the attributes in a human-readable way.
Both functions extract a derived object from the first argument using \texttt{container\_of}.
The attacker trains the call site to invoke the USB driver's method (\cref{fig:analysis-example-target})
by repeatedly displaying the attributes of that device. Next, the attacker attempts to display
the \texttt{devfreq} driver's attributes. Due to the prior training, instead of the \texttt{devfreq}
method (\cref{fig:analysis-example-actual}) being executed, the USB's method is initially speculatively
executed.
Consequently, the USB method's derived object actually points to \texttt{devfreq}'s object,
so when the USB method dereferences its \texttt{cap} field it is actually dereferencing the
value stored in the \texttt{devfreq}'s structure \texttt{max\_freq} field.  However, as shown
in~\cref{fig:analysis-example-write}, the attacker can only get \texttt{max\_freq} to contain
one of a fixed set of values.  A similar scenario, in which \texttt{max\_freq} would be some
64-bit value written by the user, would be exploitable.

\section{Discussion \& mitigations} \label{sec:discussion}

Here, we discuss possible mitigations against speculative type confusion attacks.  We distinguish
mitigations for the general problem (\cref{sec:mitigate-all}) from the specific case of eBPF (\cref{sec:mitigate-ebpf}).
We focus on immediately deployable mitigations, i.e., mainly software mitigations. Long term defenses
are discussed in~\cref{sec:related}.

\subsection{General mitigations} \label{sec:mitigate-all}

Unlike bounds check bypass gadgets, speculative type confusion gadgets
do not have a well-understood, easy to spot structure, and are difficult if not impossible for
programmers to reason about.  Mitigating them thus requires either complete Spectre protection or
statically identifying every gadget and manually protecting it.

\paragraph{Complete mitigations}
Every Spectre attack, including speculative type confusion, can be fully mitigated by placing speculation
barriers or serializing instructions after every branch. This mitigation essentially disables speculative
execution, leading to huge performance loss~\cite{sok_spectre_meltdown}.
Speculative load hardening (SLH)~\cite{SLH} (implemented in Clang/LLVM~\cite{LLVM} and ICC~\cite{ICC-mitigation})
is a more efficient complete mitigation. SLH does not disable speculative execution, but only blocks results of speculative loads
from being forwarded down the pipeline until the speculative execution proves to be correct. To this end,
SLH masks the output of every load with a mask that has a data dependency on the outcome of \emph{all} prior
branches in the program, which is obtained by emitting conditional move instructions that maintain the mask
after every branch.

\begin{figure}[t]
\includegraphics[width=\columnwidth]{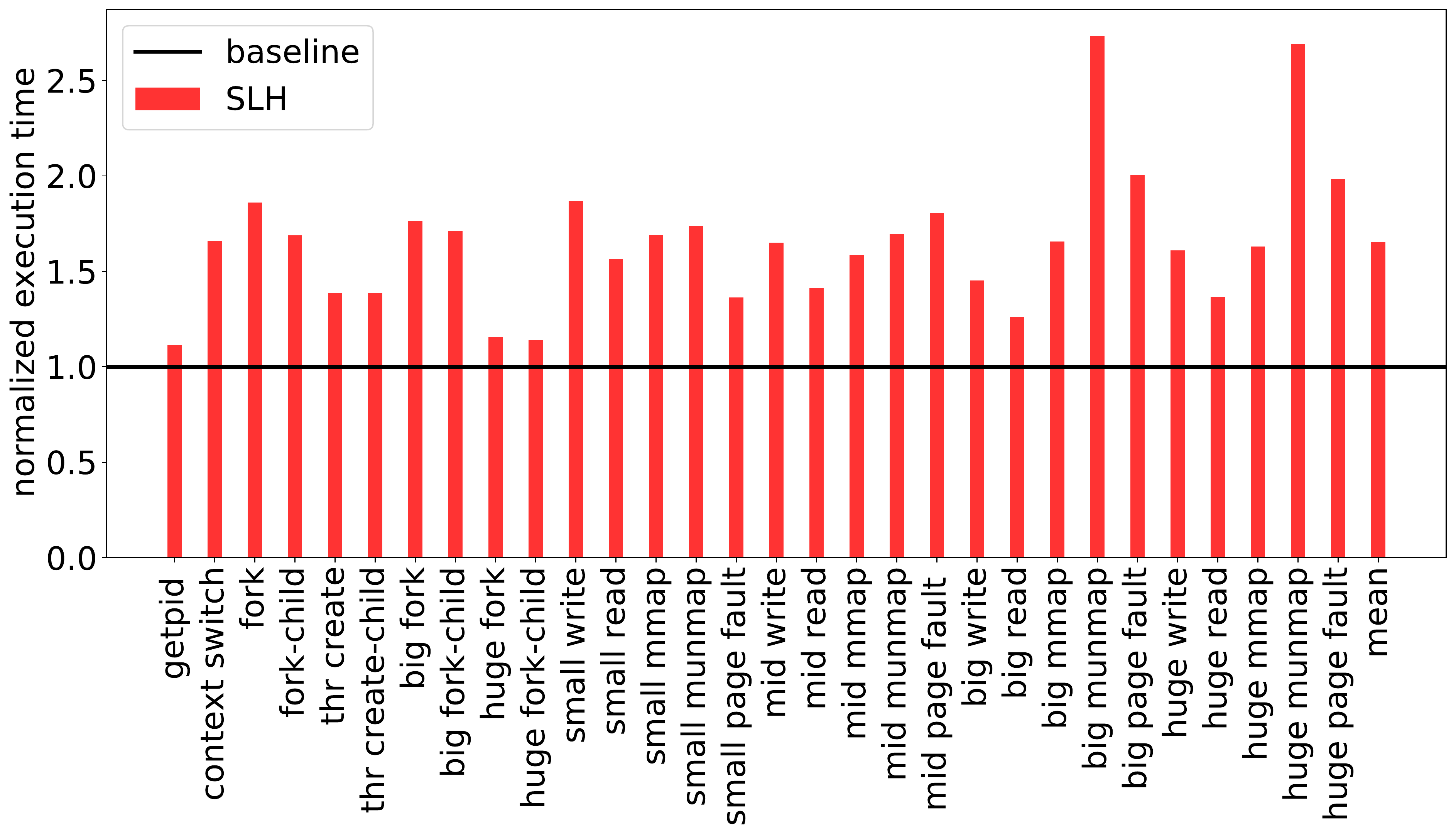}
\caption{Slowdown of Linux 5.4.119 kernel operations due to SLH.}
\label{fig:slh-linux}
\end{figure}

\begin{figure}[t]
\includegraphics[width=\columnwidth]{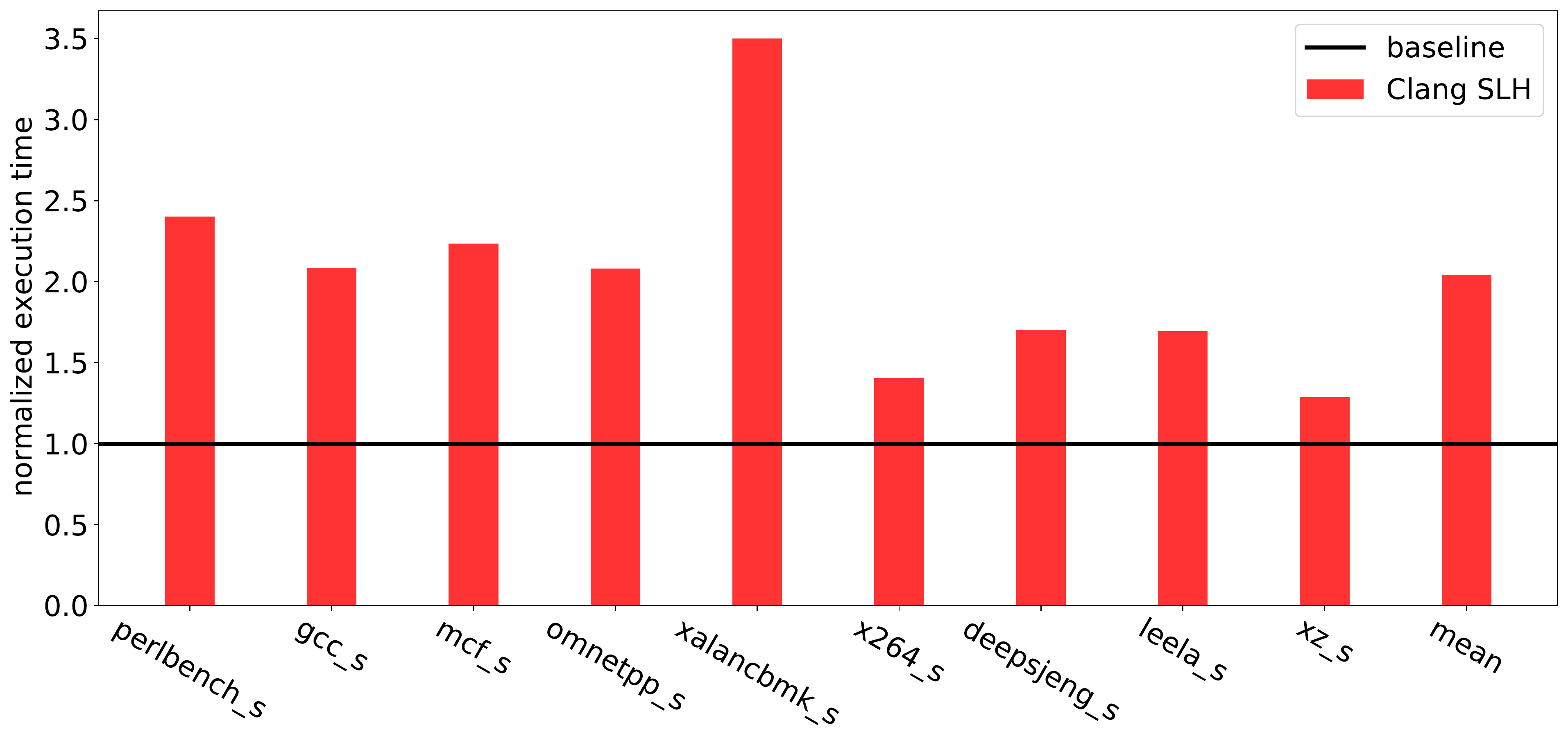}
\caption{Slowdown of SPEC CPU 2017 applications due to SLH.}
\label{fig:slh-spec}
\end{figure}

Unfortunately, we find that SLH imposes significant overhead on both kernel operations and computational workloads, far worse
than previously reported results on a Google microbenchmark suite~\cite{SLH}. To evaluate SLH's overhead on
the Linux kernel, we use LEBench~\cite{LEBench}, a microbenchmark suite that measures the performance of the
most important system calls for a range of application workloads.%
\footnote{We modify Clang/LLVM's SLH implementation to support kernel-mode execution, which is not supported
out of the box. SLH assumes that the high bits of addresses are zeroes, and
relies on this property to encode information in the stack pointer on function calls/returns~\cite{SLH}. This technique
breaks kernel execution, because the high bits of kernel virtual addresses are ones. Our modification simply
flips the bit values that SLH encodes in the stack pointer, so that non-transient kernel executions maintain a
valid stack pointer.}
To evaluate SLH's overhead on computational userspace workloads, we use the SPEC CPU2017 benchmark
suite~\cite{SPEC_CPU2017}.

We evaluate the above benchmarks on a system with a 2.6\,GHz Intel Xeon Gold 6132 (Skylake) CPU running
Linux 5.4.119.
\Cref{fig:slh-linux} shows the relative slowdown of system call execution time with an SLH-enabled
kernel, compared to a vanilla kernel (which includes Linux's standard Spectre mitigations but is compiled without SLH).
\Cref{fig:slh-spec} shows the relative execution time slowdown of a subset of the CPU2017 C/C++ benchmarks when
compiled with SLH enabled, compared to with standard compilation. In both settings, SLH imposes significant
slowdowns. SLH causes an average system call slowdown of $1.65\times$ (up to $2.7\times$) and an average CPU2017
program slowdown of $2\times$ (up to almost $3.5\times$).

Other proposed software mitigations~\cite{LAHF,Blade} use similar principles to SLH, but were evaluated on protecting
array bounds checking.
It is not clear what their overhead would be if used for complete mitigation.

\paragraph{Spot mitigations}
We contend that manual Spectre mitigation, as advocated in Linux and GCC, is not practical against
speculative type confusion.  Similarly to how transient execution attacks break the security contract between
hardware and software, speculative type confusion breaks the contract between the compiler and programs,
with correct programs possibly being compiled into vulnerable native code. Worse, any conclusion reached
about security of code can be invalidated by an unrelated code change somewhere in the program or an
update of the compiler.
Overall, human-only manual mitigation seems difficult if not infeasible.

As a result, a manual mitigation approach must be guided by a complete static analysis, which would detect every
speculative type confusion gadget in the kernel. It is notoriously difficult, however, to \emph{prove} safety of
C/C++ code, e.g., due to pointer aliasing and arithmetic~\cite{Coverity}.  Here, the problem is compounded by
the need to analyze all possible paths, which invalidates many static analysis optimizations. Indeed, current analyses
that reason about speculative execution vulnerabilities have limited scalability~\cite{Spectector}, restrict themselves to
constant-time code~\cite{ConstTimeFoundations}, or search for specific syntactic code patterns~\cite{oo7}.
Scaling an analysis to \emph{verify} that every pointer dereference in Linux
is safe from speculative type confusion is a major research challenge.

\paragraph{Hardware workarounds}
Using different BPUs for user and kernel context may be a non-intrusive hardware change that vendors can
quickly roll out.
However, this mitigation would still allow attackers to perform mistraining by
invoking in-kernel shadow branches (e.g., in eBPF programs) whose PHT entries collide with the victim's.

\subsection{Securing eBPF} \label{sec:mitigate-ebpf}

In addition to the generic mitigations, eBPF can defend from speculative type confusion in eBPF-specific
ways.  The verifier can reason about all execution flows, not just
semantically correct ones.  However, this approach would increase verification time and render
some programs infeasible to verify.  An alternative approach is for the verifier to inject masking
instructions to ensure that the operand of \emph{every} load instruction is constrained to the object
it is supposed to be accessing, generalizing the sandboxing approaches of Native Client x86-64~\cite{NACL_SFI}
and Swivel-SFI~\cite{Swivel}.

\section{Related work} \label{sec:related}

\paragraph{Attacks}
Blindside~\cite{Blindside} and SpecROP~\cite{SpecROP} employ Spectre-PHT attacks that do not involve a bounds check bypass.
Both attacks also involve indirect branching to an illegal target, whereas our exploitation of indirect branches
does not. Blindside leverages a non-speculative pointer corruption (e.g., via a buffer overflow) to speculatively hijack
control flow in the shadow of a mispredicted branch. SpecROP poisons the BTB to chain multiple Spectre gadgets with indirect
calls. With recent mitigations, SpecROP is therefore limited to intra-address space attacks and cannot target the kernel.

\paragraph{Defenses}
Non-speculative type confusion~\cite{HexType,CaVer,TypeSan,CastSan} and control-flow integrity (CFI)~\cite{CFI,KCoFI,FGCFIKS,CFIGCC,CFICOTS}
have received significant attention.  These works generally consider non-speculative memory corruption and control-flow hijacking,
not memory disclosure over covert channels.  The defenses proposed are based on the architectural semantics, and so do not straightforwardly apply to speculative execution attacks.

There are many proposals for hardware defenses against transient execution attacks.
Some designs require programmer or software support~\cite{oisa,spectre_guard,ConTExT,dawg,2018ContextSensitiveF}
but many are software-transparent. Transparent designs differ in the protection approach. Some block
only cache-based attacks~\cite{invisispec,safespec,sakalis_isca19,conditional_spec,cleanup_spec,muontrap},
whereas others comprehensively block data from reaching transient covert channels~\cite{stt,nda,SpecShield,sdo}.
These works all report drastically lower overhead than what we observe for SLH, but their results
are based on simulations.

Combining the above two lines of work, SpecCFI~\cite{SpecCFI} is a hardware mitigation for Spectre-BTB (v2)
attacks that restricts branch target speculation to legal targets, obtained by CFI analysis.
SpecCFI also assumes hardware Spectre-PHT (v1)
mitigations, and thus should not be vulnerable to speculative type confusion.

In principle, speculative type confusion can be detected by static~\cite{oo7,Spectector,ConstTimeFoundations} or dynamic~\cite{SpecFuzz}
analysis that reasons about speculative execution. To our knowledge, only {\sc Spectector}~\cite{Spectector} performs a sound analysis
targeting general-purpose code, but it has challenges scaling to large code bases, such as Linux. Other static analyses target only constant-time code~\cite{ConstTimeFoundations} or search for specific code patterns~\cite{oo7}. SpecFuzz~\cite{SpecFuzz} dynamically executes misspeculated flows, making them observable to conventional memory safety checkers, such as AddressSanitizer~\cite{asan}. Thus, SpecFuzz is not guaranteed to find all vulnerabilities.

\paragraph{eBPF}
Gershuni et al.~\cite{eBPF-StaticAnalysis} leverage abstract interpretation to design an eBPF verifier with improved precision
(fewer incorrectly rejected programs)
and scope (verifying eBPF programs with loops).
Their analysis still is based on architectural semantics, and thus does not block our described speculative type confusion
attack.

\section{Conclusion} \label{sec:conclusion}

We have shown that speculative type confusion vulnerabilities exist in the wild.
Speculative type confusion puts into question ``spot'' Spectre mitigations.
The relevant gadgets do not have a specific structure and can insidiously materialize
as a result of benign compiler optimizations and code changes, making it hard if not impossible for
programmers to reason about code and manually apply Spectre mitigations.

Speculative type confusion vulnerabilities also slip through the cracks of non-comprehensive
Spectre mitigations such as prevention of bounds check bypasses and restriction of indirect branch
targets to legal (but possibly wrong) targets.
Consequently, the Spectre mitigation approach in the Linux kernel---and possibly other systems---requires
rethinking and further research.

\section*{Disclosure}

We disclosed our findings to the Linux kernel security team, the eBPF maintainers, as well as
Google's Android and Chromium teams in June 2020. Following our report, Google awarded us a
Vulnerability Reward. The eBPF vulnerability (CVE-2021-33624) was fixed in the mainline
Linux development tree in June 2021, by extending the eBPF verifier to explore
speculative paths~\cite{ebpf-spectypeconfusion-fix}. Subsequently, we issued an advisory~\cite{our-eBPF-advisory}
to alert the various Linux distributions to the vulnerability and its mitigation.

\section*{Acknowledgements}
We thank Alla Lenchner for extending LLVM's SLH to support kernel-mode mitigation.
We thank the reviewers and our shepherd, Deian Stefan, for their insightful feedback.
This work was funded in part by an Intel Strategic Research Alliance (ISRA) grant and by
the Blavatnik ICRC at TAU.

{\footnotesize
\bibliographystyle{plain}
\bibliography{main}}

\end{document}